\documentclass[reprint,
 superscriptaddress,
 amsmath,amssymb,
 aps,
 prx
]{revtex4-2}

	\usepackage[T1]{fontenc}
	\usepackage[utf8]{inputenc}
    \usepackage[pdftex,bookmarks=false]{hyperref}
    \usepackage{microtype}
    \usepackage{graphicx}
    \usepackage{dcolumn}
    \usepackage{bm}
    \usepackage{multirow}
    \usepackage{booktabs}
    \usepackage{xcolor}
    \usepackage{mathtools}
    \usepackage{mathptmx}
    \usepackage{upgreek}
    \usepackage{float}

    \hypersetup{
    	colorlinks = true, 
		citecolor = blue,
		bookmarksnumbered = true,
		urlcolor = green
	}
	
    \DeclarePairedDelimiter\bra{\langle}{\rvert}
    \DeclarePairedDelimiter\ket{\lvert}{\rangle}
    \DeclarePairedDelimiterX\braket[2]{\langle}{\rangle}{#1 \delimsize\vert #2}

    \renewcommand{\Re}{\operatorname{Re}}
    \renewcommand{\Im}{\operatorname{Im}}
    \newcommand{\beginsupplement}{
    	\setcounter{table}{0}
    	\renewcommand{\thetable}{S\arabic{table}}
    	\setcounter{figure}{0}
    	\renewcommand{\thefigure}{S\arabic{figure}}
    }

\begin{document}

\title{Magneto-optical detection of topological contributions to the anomalous Hall effect in a kagome ferromagnet}

\date{\today}
\author{F.~Schilberth}
\altaffiliation{These two researchers share first-authorship for delivering the main experimental and theoretical contributions, respectively.}
\affiliation{Experimentalphysik V, Center for Electronic Correlations and Magnetism, Institute for Physics, Augsburg University, D-86135 Augsburg, Germany} 
\affiliation{Department of Physics, Institute of Physics, Budapest University of Technology and Economics, M\H{u}egyetem rkp. 3., H-1111 Budapest, Hungary}

\author{N.~Unglert}
\altaffiliation{These two researchers share first-authorship for delivering the main experimental and theoretical contributions, respectively.}
\affiliation{Theoretische Physik III, Center for Electronic Correlations and Magnetism, Institute for Physics, Augsburg University, D-86135 Augsburg, Germany} 

\author{L.~Prodan}
\affiliation{Experimentalphysik V, Center for Electronic Correlations and Magnetism, Institute for Physics, Augsburg University, D-86135 Augsburg, Germany}

\author{F.~Meggle}
\affiliation{Experimentalphysik II, Institute for Physics, Augsburg University, D-86135 Augsburg, Germany}

\author{J.~ Ebad Allah}
\affiliation{Experimentalphysik II, Institute for Physics, Augsburg University, D-86135 Augsburg, Germany}

\author{C.~A.~Kuntscher}
\affiliation{Experimentalphysik II, Institute for Physics, Augsburg University, D-86135 Augsburg, Germany}

\author{A.~A.~Tsirlin}
\affiliation{Experimentalphysik VI, Center for Electronic Correlations and Magnetism, Institute for Physics, Augsburg University, D-86135 Augsburg, Germany} 

\author{V.~Tsurkan}
\affiliation{Experimentalphysik V, Center for Electronic Correlations and Magnetism, Institute for Physics, Augsburg University, D-86135 Augsburg, Germany} 
\affiliation{Institute of Applied Physics, MD-2028~Chi\c{s}in\u{a}u, Republic of Moldova}

\author{J.~Deisenhofer}
\affiliation{Experimentalphysik V, Center for Electronic Correlations and Magnetism, Institute for Physics, Augsburg University, D-86135 Augsburg, Germany}

\author{L.~Chioncel}
\affiliation{Theoretische Physik III, Center for Electronic Correlations and Magnetism, Institute for Physics, Augsburg University, D-86135 Augsburg, Germany} 

\author{I.~K\'ezsm\'arki}
\affiliation{Experimentalphysik V, Center for Electronic Correlations and Magnetism, Institute for Physics, Augsburg University, D-86135 Augsburg, Germany}

\author{S.~Bord\'acs}
\affiliation{Department of Physics, Institute of Physics, Budapest University of Technology and Economics, M\H{u}egyetem rkp. 3., H-1111 Budapest, Hungary}
\email{bordacs.sandor@ttk.bme.hu}
    
\begin{abstract}
A single ferromagnetic kagome layer is predicted to realize a Chern insulator with quantized Hall conductance, which upon stacking can become a Weyl-semimetal with large anomalous Hall effect (AHE) and magneto-optical activity. Indeed, in the kagome bilayer material Fe$_3$Sn$_2$, a large AHE was detected, however, it still awaits the direct probing of the responsible band structure features by bulk sensitive methods. We measure the optical, both diagonal and Hall, conductivity spectra over a broad spectral range and identify the origin of the intrinsic AHE with the help of momentum- and band-decomposed first principles calculations. We find that low-energy transitions, tracing "helical volumes" in momentum space reminiscent of the formerly predicted helical nodal lines, substantially contribute to the AHE, which is further increased by contributions from multiple higher-energy interband transitions. Our study also reveals that local Coulomb interactions lead to band reconstructions near the Fermi level. 
\end{abstract}

\maketitle

\section{Introduction}
Recently, much effort has been focused on the study of materials derived from the kagome lattice -- the triangular lattice of corner-sharing triangles -- as they realize flat bands \cite{Xu2020, Yin2019, Mielke2021, Kang2020} and Dirac fermions \cite{Kang2020, Ye2018, Armitage2018}. Symmetry breaking phase transitions often give rise to topologically non-trivial electronic states in these compounds: magnetic Weyl-semimetals \cite{Dedkov2008, Zhang2020}, magnetic skyrmions \cite{Hou2017, Pereiro2014} and unconventional superconductivity \cite{Yin2021, Mielke2021, Ortiz2021, Ortiz2020, Uykur2021, Uykur2022}.

One of the most profound manifestations of the interplay between magnetism and the topology of the itinerant electrons is the intrinsic anomalous Hall effect (AHE) \cite{Nagaosa2010}. If the spin degeneracy of the electronic bands is lifted by breaking the time-reversal symmetry, their spin-orbit mixing leads to a finite Berry curvature, which, as a fictitious magnetic field, deflects electric currents \cite{Fang2003}. Remarkably, the AHE can become quantized in 2D by tuning the Fermi energy into the exchange gap between topologically non-trivial bands, as suggested for the case of a single kagome layer \cite{Liu2018}. When such layers are stacked in 3D, a magnetic Weyl-semimetal showing enhanced AHE response can be realized \cite{Armitage2018, Burkov2011}.

Recently, a large AHE was observed in the kagome bilayer compound Fe$_3$Sn$_2$ \cite{Wang2016, Ye2018}. The temperature independent response, assumed to represent the intrinsic AHE, was attributed to the Dirac(-like) quasiparticles emerging in the vicinity of the $K$-point. However, optical spectroscopy and STM studies also indicate the presence of flat bands close to the Fermi energy \cite{Biswas2020, Yin2018}. Therefore, the direct assignment of the bands responsible for the large intrinsic AHE is still an open issue. 

Fe$_3$Sn$_2$ (space group R$\bar3$m) consists of an alternating stack of Fe$_3$Sn kagome bilayers and honeycomb layers of Sn, as shown in the inset of Fig.~\ref{fig:measurables}(c)
\cite{Caer1978}. A ferromagnetic order develops below $T_\text{c}=657\,$K. The moments are aligned along the $c$ axis at high temperature and they gradually rotate toward the $ab$ plane as the temperature is lowered \cite{Caer1978, Hou2017, Heritage2020}. Whether the ferromagnetic order in bulk Fe$_3$Sn$_2$ is completely collinear is under debate, while in thin samples the combination of the ferromagnetic exchange, the  easy-axis anisotropy and the dipolar interactions lead to branched domains and even skyrmionic bubbles \cite{Fenner2009, Hou2017, Heritage2020, Altthaler2021}. Owing to the kagome structure, Fe$_3$Sn$_2$ shows non-trivial electronic topology. The presence of Weyl nodes in the 10\,meV vicinity of the Fermi-energy \cite{Yao2018}, several Dirac nodes at the $K$-points \cite{Ye2018, Lin2020, Biswas2020, Tanaka2020, Fang2022} as well as flat electron bands \cite{Biswas2020, Lin2018} have been proposed.

Identifying whether these features are the main source of the AHE is difficult based on magnetotransport experiments alone, since the response is a sum of the intrinsic and extrinsic contributions with sometimes identical dependence on the longitudinal conductivity \cite{Nagaosa2010}. Due to its energy-resolved nature, the off-diagonal or optical Hall conductivity spectrum $\sigma_{xy}(\omega)$ can provide the necessary information to identify the interband transitions contributing to the intrinsic AHE, as demonstrated for the nearly half-metallic CuCr$_2$Se$_4$ and Weyl semimetal candidates SrRuO$_3$ and Co$_3$Sn$_2$S$_2$ \cite{Bordacs2010, Fang2003, Shimano2011, Okamura2020}. Besides providing guidance for the assignment of inter-band transitions, the theoretical modeling based on the Density Functional Theory (DFT)~\cite{jo.gu.89,kohn.99,jone.15} is frequently used to characterize topological features~\cite{ha.ka.10} mostly based on the results of band structure calculations or associated Berry phases~\cite{xi.mi.10}.

In this paper, we report a broadband magneto-optical study of both the diagonal and Hall conductivity spectra in Fe$_3$Sn$_2$ over the energy range 50\,meV -- 2.5\,eV. Compared to ARPES, this technique probes the bulk electronic states via the transition matrix elements directly related to AHE. We observed a noticeably high $\sigma_{xy} (\omega)$ with distinct features in the infrared range, where the magneto-optical effects show a strong temperature dependence. At the low-energy cutoff of our measurements, the real part of $\sigma_{xy}(\omega)$ closely approaches the dc value of the AHE, implying that our study covers all relevant interband transitions for the intrinsic contribution. We identified three excitation continua in $\sigma_{xy} (\omega)$ yielding the main contributions to the static Hall effect. Ab initio calculations allowed us to identify the bands governing the intrinsic AHE and locate the hot spots in the Brillouin zone which dominate the diagonal and off-diagonal optical conductivity. Our study sheds light on the striking electronic features arising from the kagome units.


\section{Results and Discussion}
The broadband reflectivity and near normal incidence magneto-optical Kerr effect (MOKE) spectra were measured on the as-grown $ab$ surface of single crystals with a lateral size of $\sim$3\,mm. The latter was measured in $\pm$0.3\,T to determine the complex magneto-optical Kerr rotation, which is antisymmetric in the magnetic field. The diagonal optical conductivity spectrum was obtained by Kramers-Kronig-transformation of the reflectivity, measured over the range of 0.01 -- 2.5\,eV. The Hall conductivity spectra were calculated using the complex Kerr rotation (0.05 -- 2.5\,eV) according to:
\begin{equation}
    \theta(\omega) +i\eta(\omega)=-\frac{\sigma_{xy}(\omega)}{\sigma_{xx}(\omega)\sqrt{1+i\frac{1}{\varepsilon_0\omega}\sigma_{xx}(\omega)}},
    \label{eq:Kerr_effect}
\end{equation}
where $\omega$ is the angular frequency of the photon, $\sigma_{xx}(\omega)$ is the optical conductivity spectrum, $\epsilon_0$ is the vacuum permittivity, $\theta(\omega)$ and $\eta(\omega)$ are the Kerr rotation and ellipticity, respectively. 

\begin{figure}
    \centering
    \includegraphics[width=\columnwidth]{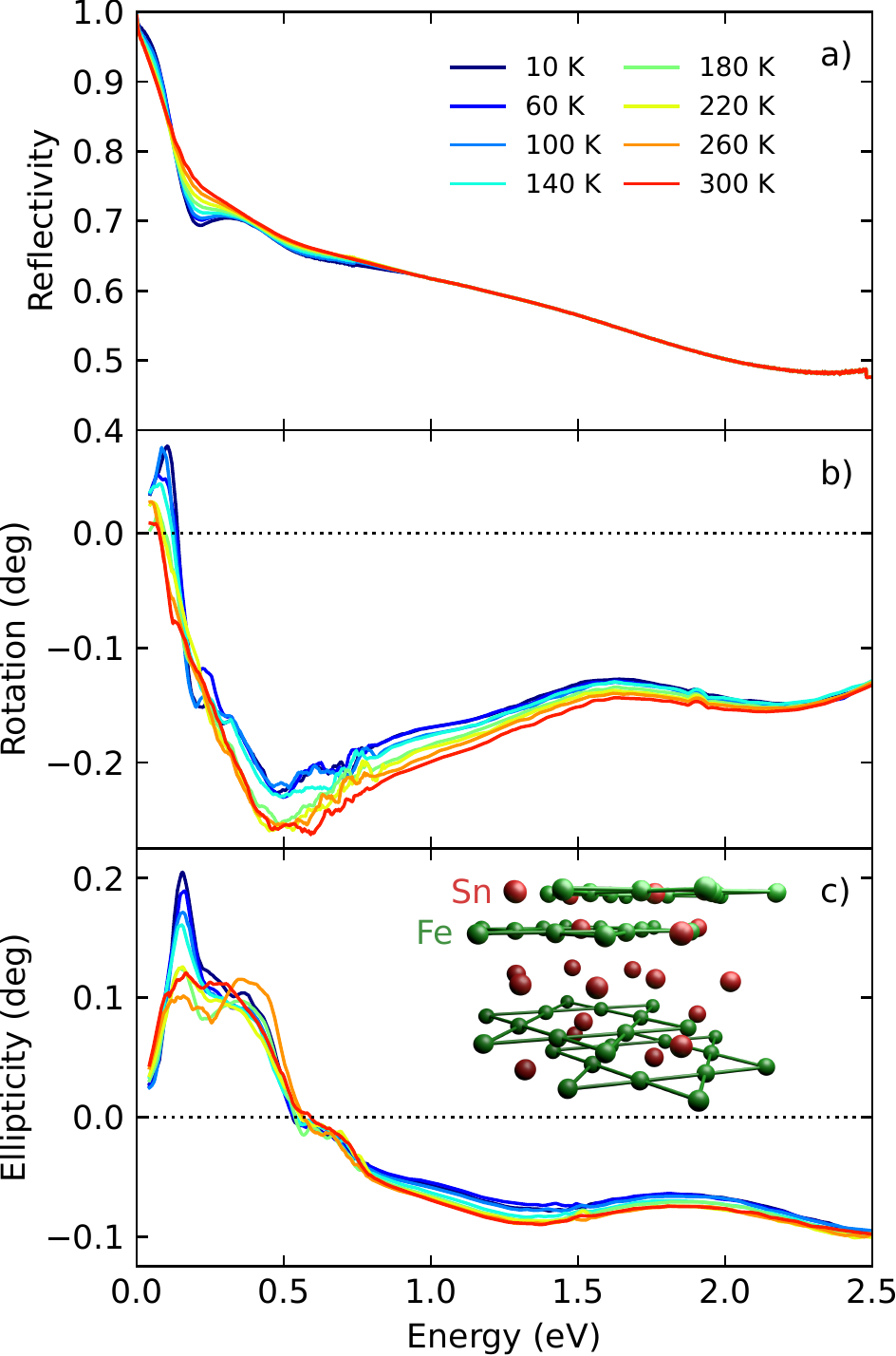}
    \caption{Spectra measured for several temperatures between 10 and 300\,K of a) the reflectivity, b) Kerr-rotation and c) ellipticity in the energy range up to 2.5\,eV. The inset in panel (c) shows a structural unit highlighting the stacking order.}
    \label{fig:measurables}
\end{figure}
\begin{figure}
	\centering
	\includegraphics[width=\linewidth]{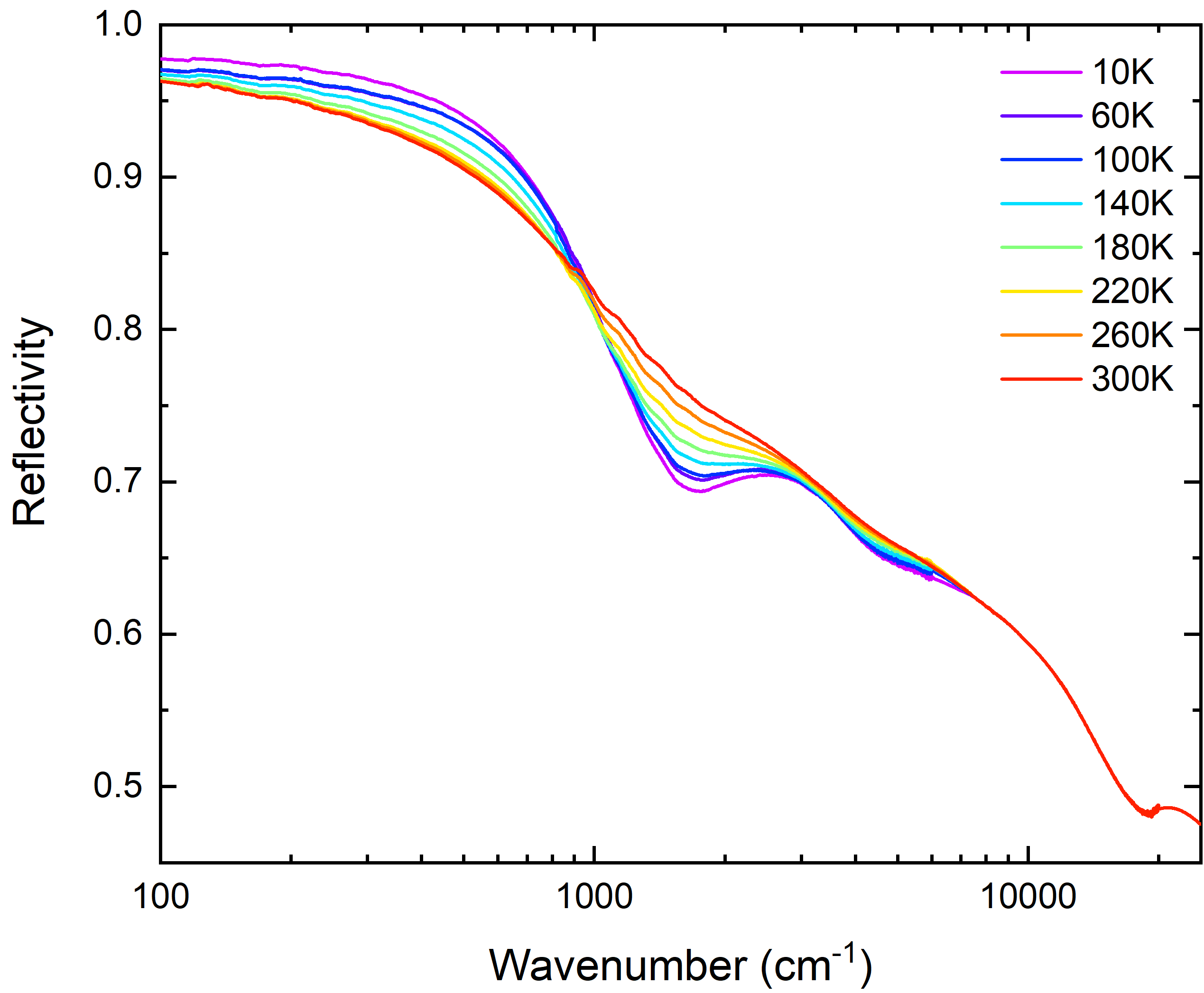}
	\caption{Temperature depedent reflectivity on log-lin-scale}   
	\label{fig:Reflectivity}
\end{figure}

We present the reflectivity and MOKE spectra measured at several temperatures from 10 to 300\,K in Fig. \ref{fig:measurables}. In agreement with former results \cite{Biswas2020}, the reflectivity shows metallic behaviour (For log-lin scale compare Fig.~\ref{fig:Reflectivity}): towards zero frequency it approaches unity and at 0.125\,eV, it drops corresponding to the plasma edge, that becomes sharper toward low temperatures. Since the magnetic order in Fe$_3$Sn$_2$ develops well above the studied temperature range, we detected a finite Kerr effect already at room temperature. The simultaneously measured rotation and ellipticity spectra can be closely mapped on each other by Kramers-Kronig-transformation, validating the measurement procedure. Accompanying the plasma edge, the rotation spectra show a steep increase whereas the ellipticity has a peak which becomes more pronounced toward low temperatures. The dielectric response appearing in the denominator in Eq.~\ref{eq:Kerr_effect} is strongly suppressed at the plasma frequency and correspondingly enhances the MOKE \cite{Feil1987}. Around 0.5\,eV, the rotation exhibits a broad minimum and the ellipticity shows a zero-crossing. At higher energies, both Kerr parameters remain negative with a small monotonous temperature dependence. Above 0.8\,eV, the overlapping interband transitions form a rather featureless optical response \footnote{The experimental details together with the dc AHE data and details of the crystal growth are presented in the Supplemental Material.}.

\begin{figure}
    \centering
    \includegraphics[width=\linewidth]{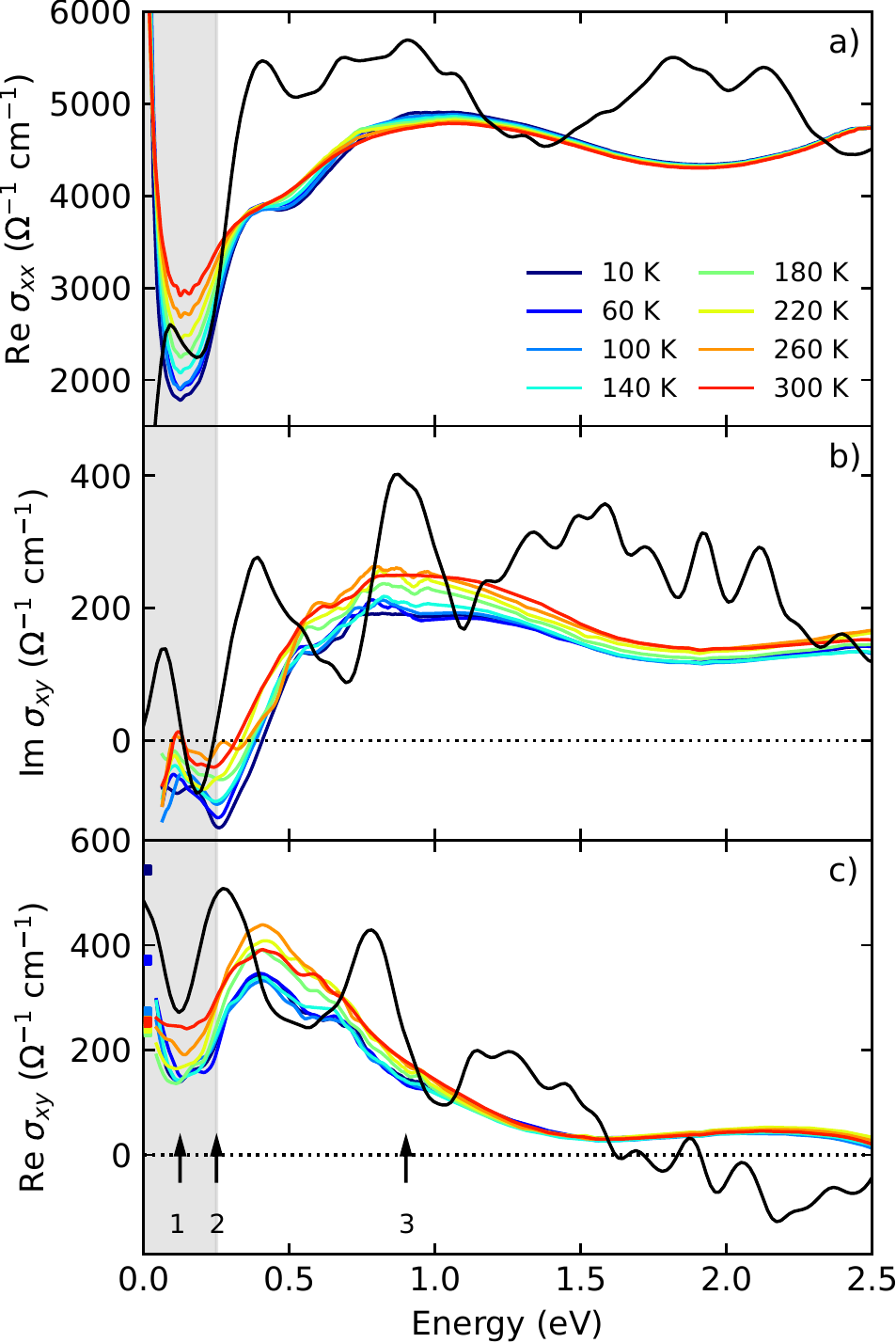}
    \caption{Comparison of the experimental conductivity spectra measured between 10 and 300\,K (colored lines) and the theoretical DFT spectra (black lines) calculated as described in the text. a), b) and c) respectively show the real part of the diagonal, $\Re\,\sigma_{xx}$, as well as, the imaginary and real part of the off-diagonal conductivity spectra, $\Im\,\sigma_{xy}$ and $\Re\,\sigma_{xy}$. For $\Re\,\sigma_{xy}$, the static AHE values are plotted for comparison.}
    \label{fig:exp_theory}
\end{figure}
\begin{figure*}
	\centering
	\includegraphics[width=\linewidth]{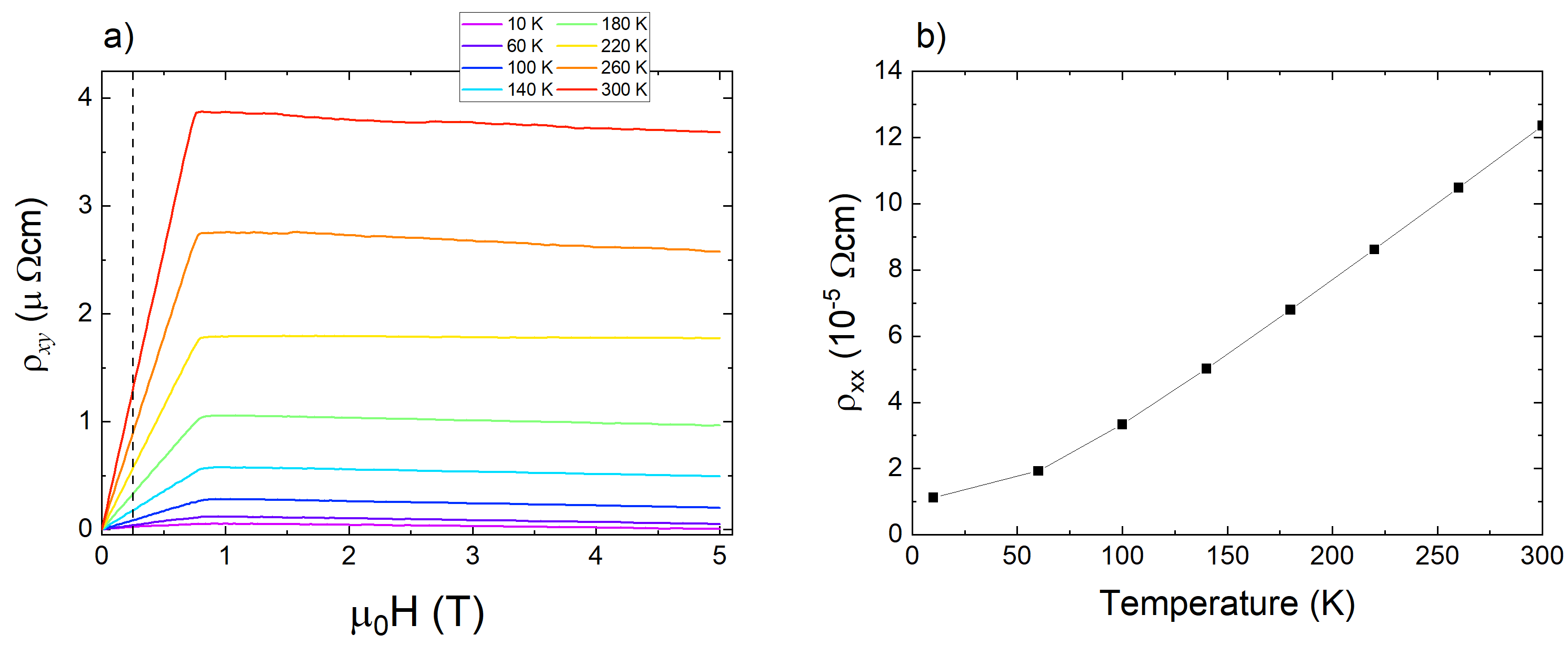}
	\caption{a) Magnetic field dependent dc-Hall resistivity measured at different temperatures. The dashed line indicates the measurement field for the MOKE-experiments. b) Temperature dependent dc resistivity measurement}   
	\label{fig:dc_Hall}
\end{figure*}

  
The derived diagonal and the off-diagonal elements of the conductivity tensor are displayed in Fig.~\ref{fig:exp_theory}. In the real part of the diagonal conductivity, $\Re\,\sigma_{xx}$, there is a step edge at 0.25\,eV and a broad hump centered at 0.9\,eV, both located well above the Drude tail. Our diagonal optical conductivity spectrum agrees with that published in Ref.~\onlinecite{Biswas2020}. Fig.~\ref{fig:exp_theory}(b) displays the imaginary part of the off-diagonal conductivity corresponding to the absorption difference for left and right circularly polarized photons. $\Im\,\sigma_{xy} (\omega)$ shows three dominant features: a small positive peak at 0.1\,eV (1), a minimum at 0.25\,eV (2) and a broad hump around 0.9\,eV (3), their energies indicated by arrows. The minimum slightly shifts to higher energies toward low temperatures, whereas the magnitude of $\Im\,\sigma_{xy}$ at higher energies decreases. As expected from the Kramers-Kronig connection between Re and $\Im\,\sigma_{xy}$, the derivative shape of these features appears in $\Re\,\sigma_{xy}$ as shown in Fig.~\ref{fig:exp_theory}(c). For comparison, the dc Hall conductivity values extracted from the measurement in Fig.~\ref{fig:dc_Hall} are included also.

Importantly, at the low-energy cutoff, 0.05\,eV, the $\Re\,\sigma_{xy}$ spectra closely match the corresponding static values, except for temperatures below 100\,K, where the increase of the dc AHE is suspected to be caused by enhanced extrinsic contributions \cite{Ye2018, Wang2016}. This implies that $\Re\,\sigma_{xy}$ extends smoothly to the dc limit. In other words, our data mainly captures the interband transitions, which govern the intrinsic dc AHE.


To elucidate the origin of the spectral features, we performed DFT calculations in \textsc{wien2k}~\cite{Blaha2021} using a GGA functional~\cite{pe.bu.96} for the exchange correlation energy and with spin-orbit coupling introduced in a second variational procedure. Although shown and discussed in several previous publications~\cite{Lin2018, Lin2020, Tanaka2020, Yao2018} we consider first in some detail the electronic structure and the magnetic properties of Fe$_3$Sn$_2$. It is well known that DFT (GGA) fails to properly describe the ground state of many systems including relatively narrow $d$-states. In particular, we found that applying the $+U$ corrections significantly improves the agreement between the calculated and measured optical conductivity spectra in Fe$_3$Sn$_2$. We considered the so-called Hubbard correction $U=1.3$\,eV to Fe($3d$)-orbitals only, and set $J=0$ for the Hund parameter to ensure that the Fe spin and orbital moments in the ferromagnetic ground state remain of the same magnitude as for $U=0$, $\mu_\mathrm{Fe,spin}^\mathrm{GGA(+U)+SO} = 2.5\,\mu_\mathrm{B}$ and $\mu_\mathrm{Fe,orb.}^\mathrm{GGA(+U)+SO} = 0.1\,\mu_\mathrm{B}$, respectively. Note that the same U and J values provide a good agreement between the theoretical and experimental ARPES spectra~\footnote{https://www.youtube.com/watch?v=dWJxK6io0UA, in private communication, the presenter confirmed that these results have entered the publication process. Once available the reference will be updated.}
and is in the range of values used in a different publication~\cite{Yao2018}.

\begin{figure}
    \centering
\includegraphics[width=\columnwidth]{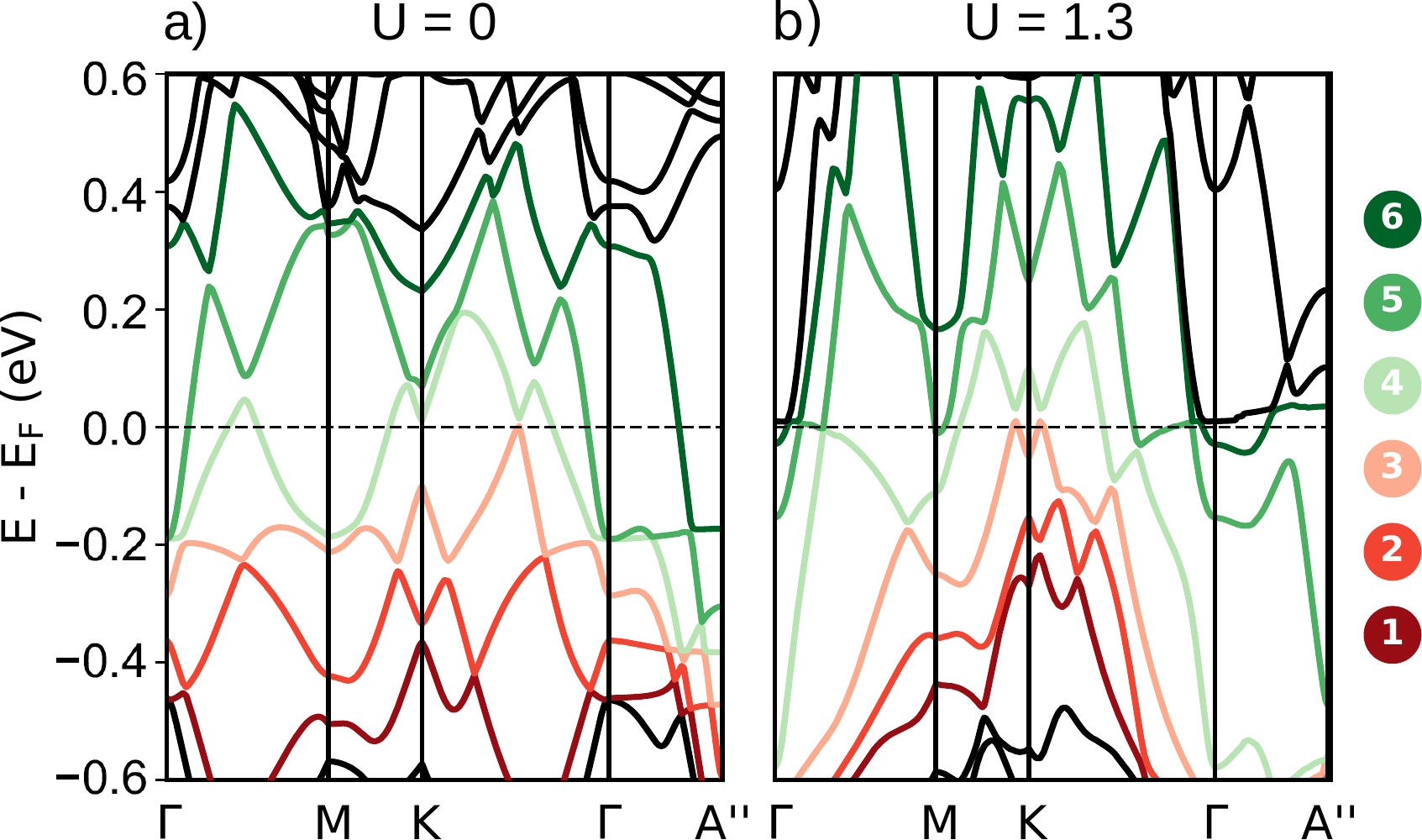}
    \caption{Panel a) and b) show the band structure of Fe$_3$Sn$_2$ as determined using GGA+SO without and with +U correction, respectively. Bands relevant for the low energy transitions are assigned by indices ranging from 1 to 6 and are colored correspondingly. For the assignment of the high-symmetry points refer to Fig.~\ref{fig:energy_diff}.}
    \label{fig:bands}
\end{figure}

The band structure calculated for $U$=0, displayed in Fig.~\ref{fig:bands}(a), is in a good agreement with the results reported by Ref.~\onlinecite{Fang2022}, where two helical nodal lines in close vicinity to the $H'-K-H''$ line are identified. The emergence of these nodal lines was attributed to the trigonal stacking of kagome layers and to the rhombohedral symmetry in Fe$_3$Sn$_2$. As expected, the effect of the Hubbard $U$ parameter leads to a re-organization of the energy bands as shown in  Fig.~\ref{fig:bands}(b). We therefore observed a disruption of the nodal lines (derived from the DFT Hamiltonian computation of Ref.~\onlinecite{Fang2022}) due to the presence of effective Hubbard interactions. The double "almost touching" bands (see bands 3 and 4 around the $K$ point in Fig.~\ref{fig:bands}) are pushed towards the Fermi edge, making them accessible for low-energy optical transitions. 

\begin{figure*}
    \centering
    \includegraphics[width=1.0\textwidth]{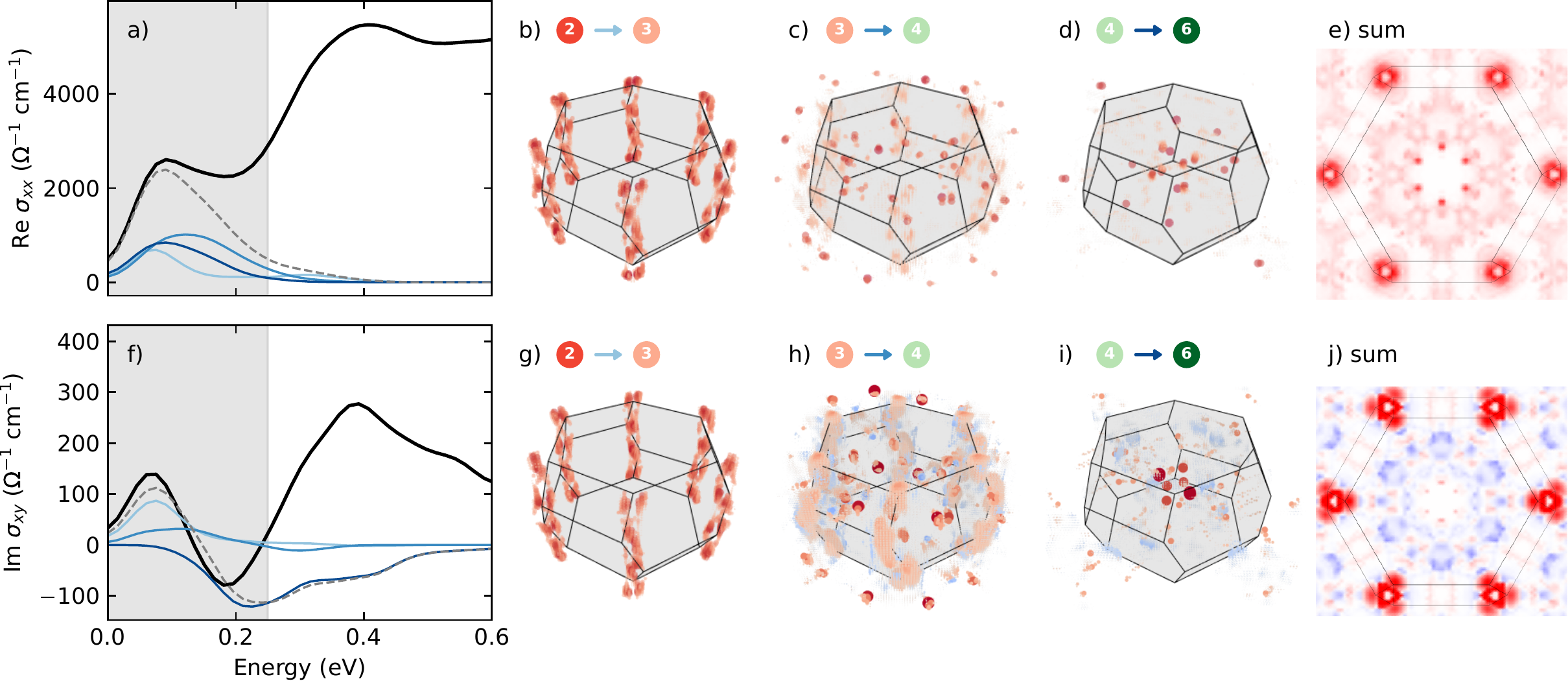}
    \caption{Decomposition of the calculated spectra into the three transitions ($2\rightarrow 3$, $3\rightarrow 4$ and $4\rightarrow 6$) with maximum spectral weight for the energy window of 0 -- 0.25\,eV (shaded in grey). (a)/(f) The total spectrum calculated for $\Re\,\sigma_{xx}$/$\Im\,\sigma_{xy}$ is shown in black, while contributions of the three transitions are plotted in blue shades and their sum in dashed grey. (b)-(d) $k$-resolved spectral weight distributions $\rho_{\alpha \beta}^{n\rightarrow n^\prime}(\bm{k},\,0\,\text{eV}<\hbar\omega<0.25\,\text{eV})$ of the three transitions and their sum (e) for $\Re\,\sigma_{xx}$. The numerical labels refer to the bands as in Fig.~\ref{fig:bands} while the arrows have the same color as the corresponding spectra in panel (a). Panels (g)-(j) depict the same information for $\Im\,\sigma_{xy}$, where $k$-regions with red/blue color represent positive/negative values. As seen in panels (b) and (g), the lowest energy transitions are located around the $H'-K-H''$ line as ''helical volumes'' and dominate the off-diagonal response around 0.1\,eV.}
    \label{fig:decomp}
\end{figure*}


We calculate the dissipative part of optical spectra, i.e.~$\Re\,\sigma_{xx}$ and $\Im\,\sigma_{xy}$ using the Kubo linear response theory~\cite{kubo.57}.
\begin{equation}\label{eq:sigma}
\begin{aligned}
    \sigma_{\alpha\beta}(\omega) &=
		\frac{4 \pi e^2}{m^2\omega^2} \sum_{n,n^\prime} \int d \bm{k}
		\bra{n\bm{k}} \hat{p}_\alpha \ket{n^\prime\bm{k}} \bra{n^\prime\bm{k}} \hat{p}_\beta \ket{n\bm{k}} \\
	&\cdot f_{n{\mathbf{k}}} (1-f_{n^\prime{\mathbf{k}}})	 \delta \big( \epsilon_{n \bm{k}} - \epsilon_{n^\prime \bm{k}} - \omega \big)
\end{aligned}
\end{equation}
with $\alpha,\beta=x,y,z$ and $\hat{p}_{\alpha/\beta}$ being the components of the momentum operator; $f_{n{\mathbf{k}}}$ is the Fermi function.  Eq.~\ref{eq:sigma} contains the sum over the bands (the band index also includes spin) where only interband ($n\ne n^\prime$) contributions are considered. The theoretical spectrum of the $\Re\,\sigma_{xy}$ was obtained by Kramers-Kronig transformation of the imaginary part. For plotting, the spectra are convoluted with a gaussian with a FWHM of 50\,meV.

\begin{figure}
	\centering
	\includegraphics[width=\linewidth]{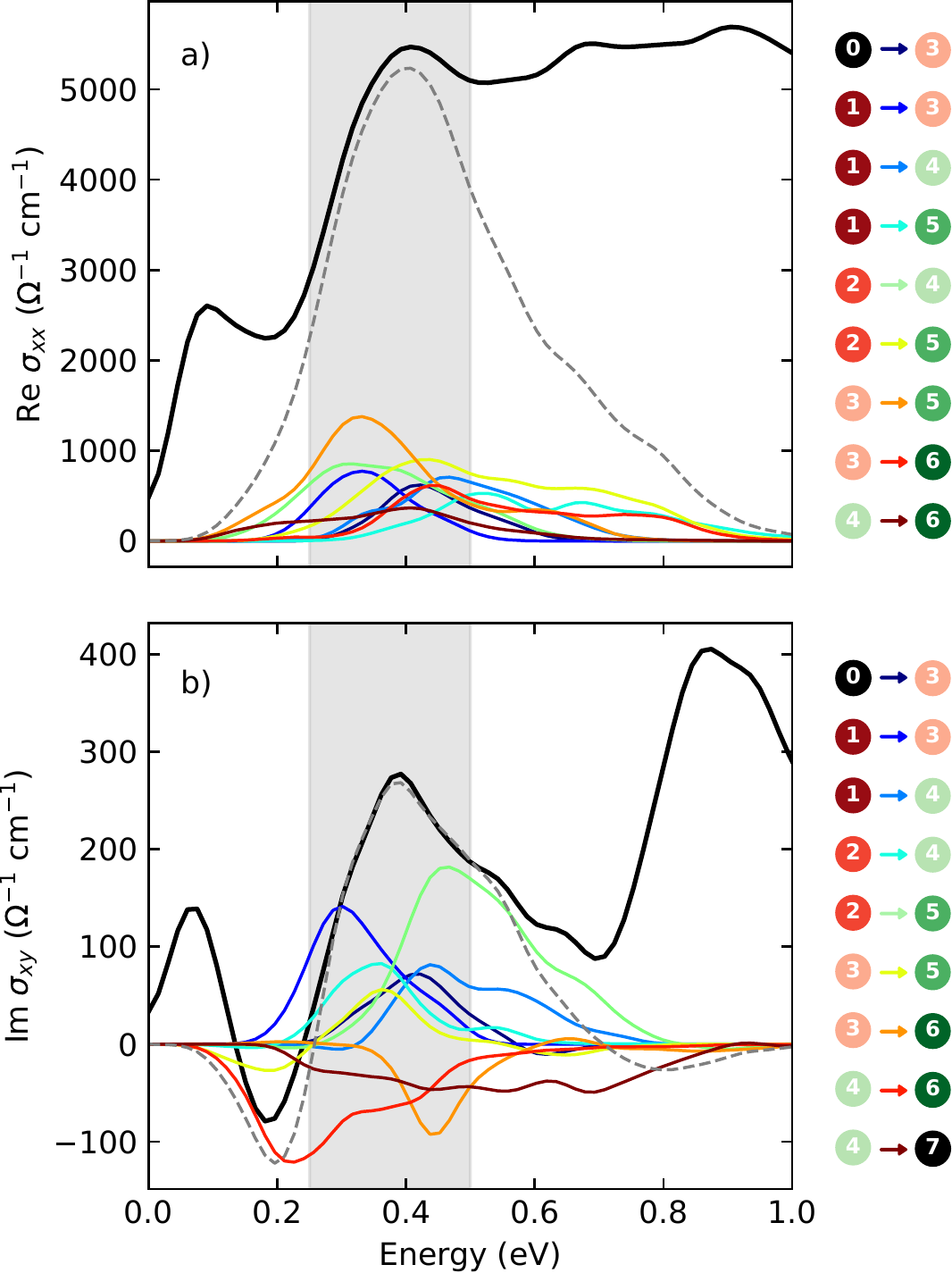}
	\caption{Spectral decomposition for the energy window from 0.25-0.5\,eV. The bands 0 and 7 not indicated in the bandstructure lie just below band 1 and above band 6, respectively.}   
	\label{fig:step_decomp}
\end{figure}

We present the overall comparison of the theoretical and experimental spectra in Fig.~\ref{fig:exp_theory}. Apart from the intraband (Drude) contributions that are not included in the calculations, the theory correctly captures the low-energy diagonal optical conductivity $\Re\,\sigma_{xx}$. Importantly, the calculations reproduce the prominent step edge at 0.25\,eV, and they indicate that multiple bands contribute significantly to this feature (see Fig.~\ref{fig:step_decomp}). The calculated imaginary and real part of $\sigma_{xy}$ in panels b and c, respectively, reveal more spectral features that are also in good agreement with the experiments. In $\Im\,\sigma_{xy}$, the theory resolves a positive peak at 0.1\,eV, a minimum at 0.2\,eV and another positive peak at 0.4\,eV that are all present in the experimental data though slightly shifted to higher energies. The peak occurring around 0.9\,eV appears in the measurement as a broad plateau centered at the same energy which swallows the surrounding sharper features. Since the theory curves are sharper than the experimental spectra, a larger frequency independent broadening could improve the overall correspondence, but the low frequency features would certainly be masked. Although only interband transitions are included in the calculations, the dc extrapolation of theoretical $\Re\,\sigma_{xy}$ agrees well with the range of measured AHE, which implies largely intrinsic AHE in Fe$_3$Sn$_2$. We emphasize that none of the individual features in $\Im\,\sigma_{xy}$ discussed above is responsible for the static Hall-effect alone, but their interplay can fully capture the dc AHE.


Having established the connection between theoretical and experimental spectra, we now investigate the origin of the optical response by band- and $k$-resolved calculations. It is common practice, to split the band summations in Eq.~\ref{eq:sigma} and examine the contributions of individual transitions. Especially for simple systems with a clearly arranged band manifold, this is often sufficient to assign certain observed transitions to features in the electronic structure. However, for complex multiband systems it turns out to be cumbersome to apply this type of analysis. Furthermore, one tends to miss important contributions, arising from regions in reciprocal space which lie off the high symmetry lines usually employed to visualize the electronic band structure. Therefore, to disentangle the contributions arising from the multitude of bands in Fe$_3$Sn$_2$, we performed the analysis of the optical spectra in different photon energy windows, monitoring the spectral density $\rho_{\alpha \beta}^{n\rightarrow n^\prime}(\bm{k}, \omega)$-decomposition of the optical conductivity: 
$        \sigma_{\alpha \beta}(\omega) = 
        \int_{BZ} d \bm{k}  \rho_{\alpha \beta}^{n\rightarrow n^\prime}(\bm{k}, \omega) 
$.
We determined the most important interband transitions by means of their spectral weight within the chosen energy window.

In Fig.~\ref{fig:decomp}, we present the results of this analysis for $\Re\,\sigma_{xx}$ and $\Im\,\sigma_{xy}$ in the photon energy window $\hbar\omega=0 - 0.25$\,eV, corresponding to experimental features 1 and 2 in Fig.~\ref{fig:exp_theory}. Three transitions (between the bands $2\rightarrow 3$, $3\rightarrow 4$ and $4\rightarrow 6$) reproduce the total spectra within this energy window almost perfectly. Looking at the individual transitions in Fig.~\ref{fig:decomp}(b) – (e) reveals that the low energy peak in $\Re\,\sigma_{xx}$ is partially formed by transitions close to the $K$ point ($2 \rightarrow 3$, $3 \rightarrow 4$), i.e. hot spots of the spectral density trace a double helix in k-space. This resembles the shape of two gapped helical nodal lines formed by bands $2$ and $3$ and travelling along the $H'-K-H''$ line, similar to the structures found in Ref.~\onlinecite{Fang2022}. For direct observation, Fig.~\ref{fig:energy_diff} in the Supplement compares the spectral density between bands 2 and 3 with their eigenvalue difference, yielding similar helical shapes.

The $\Im\,\sigma_{xy}$ spectrum is analyzed in a similar fashion in Fig.~\ref{fig:decomp}(f) - (j)). Note, that for $\sigma_{xy}$ negative contributions (colored blue in the Brillouin zone plots) can occur in the spectral density, arising from negative momentum matrix element products. The peak at 0.1\,eV arises dominantly from transitions between bands $2 \rightarrow 3$ with the optical weight located around the $K$-point in the same helical fashion as for $\Re\,\sigma_{xx}$. By contrast, the dip around 0.2\,eV results from transition located in the bulk of the Brillouin zone between bands $4 \rightarrow 6$ with dominantly negative spectral density. 

This analysis demonstrates the strength of magneto-optical spectroscopy in the investigation of low energy topological electronic structures. While in the experimental $\Re\,\sigma_{xx}$ spectra, the low-energy peak (feature 1) is masked by the large Drude contribution, the low energy interband transitions are clearly distinguished in $\sigma_{xy}$ which enables direct experimental observation and grants valuable information way beyond the insights obtained from standard optical spectroscopy.

\section{Summary}

In summary, our magneto-optical study of the low-energy excitations allowed us to identify the electronic structures responsible for the AHE in Fe$_3$Sn$_2$. In the optical Hall conductivity, we detected a series of spectral features all of which give substantial contribution to the static AHE. We found that their signs are alternating i.e.~their contributions to the dc AHE compete with each other. These interband transitions fully determine the AHE at 100\,K and above where the intrinsic AHE is dominant. In agreement with Ref.~\onlinecite{Fang2022}, our band and k-resolved calculation indicate helical electronic states emerging from the interplay of the spin-orbit coupling essential for topological semimetals and the inherent trigonal stacking of the kagome layers in Fe$_3$Sn$_2$. We found significant contributions to the low energy $\sigma_{xx}$ and $\sigma_{xy}$ within these ''helical-volumes'' in momentum space. Electronic states situated within this volume may form ``helical'' electronic liquids in the presence of electronic interactions, that are found to be important in the present case. In analogy to particle physics, low energy Hamiltonians can mimic Dirac fermions with a defined helicity according to their momentum and spin. Consequently, also the helicity can serve as a quantum number at the Fermi level. While the spin degeneracy may be lifted by an external magnetic field, the degeneracy from helicity as a time-reversal invariant is preserved resulting in a helical liquid. However, if this degeneracy is lifted in an interacting multi-orbital system, with the Fermi energy intersecting only bands of one helicity, than the conduction modes will be formed by helical states whose spin is fixed by the propagation direction. The tuneability of these states may be achieved by electric and/or magnetic fields making them highly relevant for spintronic applications.

Finally, we note that our magneto-optical approach is widely applicable for other materials, where the interplay of magnetism and electronic topology leads to a large AHE. Since these bulk measurements are sensitive to the position of the Fermi energy, they are also suitable to evaluate the effect of doping, which can serve as a tool to control the Hall response.

\begin{acknowledgements}
    This research was partly funded by Deutsche Forschungsgemeinschaft DFG via the Transregional Collaborative Research Center TRR 80 ''From Electronic correlations to functionality'' (Augsburg, Munich, Stuttgart). This work was supported by the Hungarian National Research, Development and Innovation Office – NKFIH grants FK 135003 and Bolyai 00318/20/11 and by the Ministry of Innovation and Technology and the National Research, Development and Innovation Office within the Quantum Information National Laboratory of Hungary. Sándor Bordács is supported by the ÚNKP-21-5-BME-346 new national excellence program of the ministry for innovation and technology from the source of the national research, development and innovation fund.
\end{acknowledgements}

\clearpage
\beginsupplement
\section*{Supplemental Material}
\subsection*{Crystal growth}
Fe$_3$Sn$_2$ single crystals have been synthesized by the chemical transport reactions method. As starting material, preliminary synthesized polycrystalline powder prepared by solid state reactions from the high-purity elements was used with iodine as transport agent. The growth was performed in quartz ampoules in two-zone furnaces at temperatures between 730 and 680\,$^\circ$C. After 4 -- 6 weeks of transport, the plate-like single-crystalline samples of thickness 20 -- 40\,$\upmu$m along the $c$-axis and 3 -- 5\,mm within the $ab$-plane were found in the hot part of ampoule.

\subsection*{Reflectivity measurements}
The reflectivity spectra of Fe$_3$Sn$_2$ were obtained by using a Hyperion IR-microscope, equipped with a 15x Cassegrian objective, coupled to a Bruker Vertex 80v FT-IR spectrometer. The spectra were measured in the frequency range 120-6000~cm$^{-1}$ (0.01-0.74\,eV) from room temperature down to 7 K. A silver film, evaporated on half of the measured surface of the Fe$_3$Sn$_2$ single crystal was used as a reference. Each low frequency spectrum was merged to the NIR-VIS spectrum [6000 to 20000~cm$^{-1}$ (0.74-2.5~eV)] which was measured at room temperature. The optical conductivity $\sigma_1$ was calculated by using Kramers-Kronig analysis. At this point, the low energy of the reflectivity spectrum (0-120~cm$^-1$) was extrapolated by using a Drude-Lorentz fitting, while above 2.5~eV the reflectivity spectrum was extrapolated to high energies using x-ray atomic scattering functions \cite{Tanner.2015}. 

\subsection*{MOKE-spectroscopy}
The broadband MOKE spectra were recorded in near-normal incidence and were combined from several measurements in different frequency ranges, employing grating and interferometer based spectrometers as described elsewhere \cite{Sato1981, Demko2012, Bordacs2010}. Small permanent magnets provided a field of 0.3\,T at the sample position. For the FIR-MOKE experiments, two polarisers with high extinction-ratio were placed at $45^\circ$ relative orientation before and after the sample in a FTIR-spectrometer. In this setting, neglecting all nonlinear terms, the intensity after the analyser is directly proportional to the Kerr-rotation. For calibration, the analyser was rotated to $46^\circ$, yielding the intensity change of a $1^\circ$ rotation. Same as for the polarisation-modulation based technique, two spectra for positive and negative static magnetic fields were antisymmetrised to obtain the signal odd in magnetic field. The corresponding ellipticity spectra were obtained by simultaneously fitting the merged rotation spectra and the higher frequency ellipticity spectra in Reffit \cite{Kuzmenko2005}.

\subsection*{DFT-calculations}
We performed these within Density Functional Theory using the \textsc{wien2k}~\cite{Blaha2021} code. The generalized gradient approximation as parameterized by Perdew-Burke-Ernzerhof was used as exchange-correlation functional. Spin-orbit interaction was introduced in a second variational procedure. 1000 $k$ points were used to sample the full Brillouin zone in the self-consistent calculation. A $rkmax$ value of 8 was chosen as cutoff for the plane wave expansion. The eigenvalues and momentum matrix elements for the evaluation of the optical response were evaluated on a larger grid of 10000 $k$ points. Momentum matrix elements were calculated using the built-in \textit{optics} module \cite{ambrosch-draxl_linear_2006}. The crystal structure was taken from Fenner et al. \onlinecite{Fenner2009} with the magnetization pointing along the $c$-axis.

\begin{figure*}
	\centering
	\includegraphics[width=\linewidth]{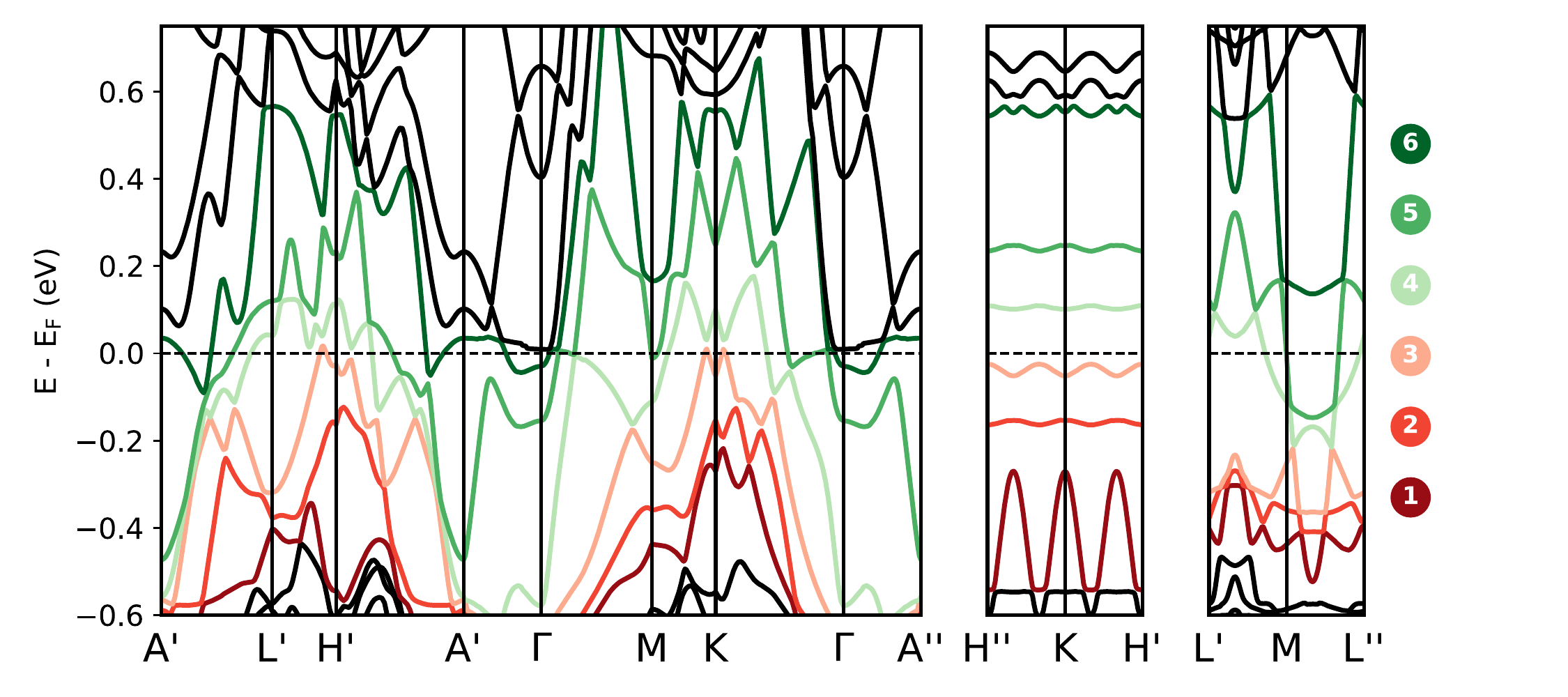}
	\caption{Full band structure for $U=1.3$\,eV.}   
	\label{fig:band-structure}
\end{figure*}

\begin{figure*}
	\centering
	\includegraphics[width=\linewidth]{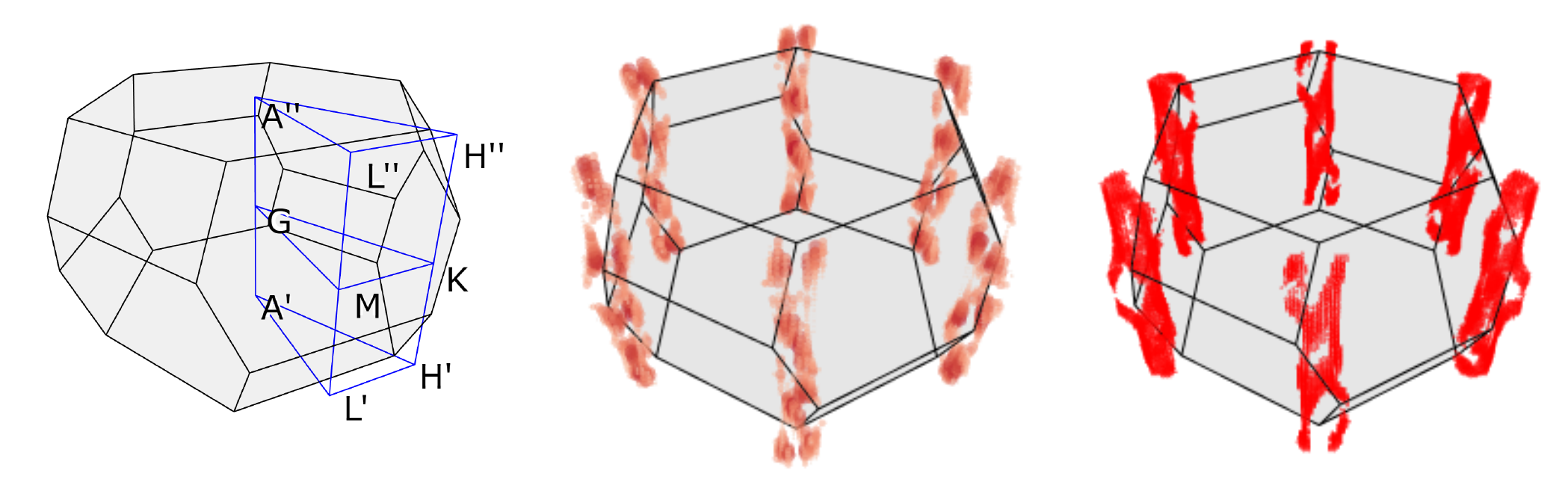}
	\caption{Left: Brillouin zone with hexagonal slab and the corresponding high-symmetry points. Center: double helical feature seen in $\Re\,\sigma_{xy}$ for transition $2 \rightarrow 3$ in the 0 -- 0.25\,eV energy window. Right: Plot of $k$-points, where the eigenvalue difference between bands 2 and 3 is below a threshold of 50\,meV.}   
	\label{fig:energy_diff}
\end{figure*}

\begin{thebibliography}{45}%
	\makeatletter
	\providecommand \@ifxundefined [1]{%
		\@ifx{#1\undefined}
	}%
	\providecommand \@ifnum [1]{%
		\ifnum #1\expandafter \@firstoftwo
		\else \expandafter \@secondoftwo
		\fi
	}%
	\providecommand \@ifx [1]{%
		\ifx #1\expandafter \@firstoftwo
		\else \expandafter \@secondoftwo
		\fi
	}%
	\providecommand \natexlab [1]{#1}%
	\providecommand \enquote  [1]{``#1''}%
	\providecommand \bibnamefont  [1]{#1}%
	\providecommand \bibfnamefont [1]{#1}%
	\providecommand \citenamefont [1]{#1}%
	\providecommand \href@noop [0]{\@secondoftwo}%
	\providecommand \href [0]{\begingroup \@sanitize@url \@href}%
	\providecommand \@href[1]{\@@startlink{#1}\@@href}%
	\providecommand \@@href[1]{\endgroup#1\@@endlink}%
	\providecommand \@sanitize@url [0]{\catcode `\\12\catcode `\$12\catcode
		`\&12\catcode `\#12\catcode `\^12\catcode `\_12\catcode `\%12\relax}%
	\providecommand \@@startlink[1]{}%
	\providecommand \@@endlink[0]{}%
	\providecommand \url  [0]{\begingroup\@sanitize@url \@url }%
	\providecommand \@url [1]{\endgroup\@href {#1}{\urlprefix }}%
	\providecommand \urlprefix  [0]{URL }%
	\providecommand \Eprint [0]{\href }%
	\providecommand \doibase [0]{https://doi.org/}%
	\providecommand \selectlanguage [0]{\@gobble}%
	\providecommand \bibinfo  [0]{\@secondoftwo}%
	\providecommand \bibfield  [0]{\@secondoftwo}%
	\providecommand \translation [1]{[#1]}%
	\providecommand \BibitemOpen [0]{}%
	\providecommand \bibitemStop [0]{}%
	\providecommand \bibitemNoStop [0]{.\EOS\space}%
	\providecommand \EOS [0]{\spacefactor3000\relax}%
	\providecommand \BibitemShut  [1]{\csname bibitem#1\endcsname}%
	\let\auto@bib@innerbib\@empty
	\bibitem [{\citenamefont {Xu}\ \emph {et~al.}(2020)\citenamefont {Xu},
		\citenamefont {Zhao}, \citenamefont {Yi}, \citenamefont {Wang}, \citenamefont
		{Yin}, \citenamefont {Wang}, \citenamefont {Hu}, \citenamefont {Wang},
		\citenamefont {Liu}, \citenamefont {Xu}, \citenamefont {Lu}, \citenamefont
		{Soluyanov}, \citenamefont {Lei}, \citenamefont {Shi}, \citenamefont {Luo},\
		and\ \citenamefont {Chen}}]{Xu2020}%
	\BibitemOpen
	\bibfield  {author} {\bibinfo {author} {\bibfnamefont {Y.}~\bibnamefont
			{Xu}}, \bibinfo {author} {\bibfnamefont {J.}~\bibnamefont {Zhao}}, \bibinfo
		{author} {\bibfnamefont {C.}~\bibnamefont {Yi}}, \bibinfo {author}
		{\bibfnamefont {Q.}~\bibnamefont {Wang}}, \bibinfo {author} {\bibfnamefont
			{Q.}~\bibnamefont {Yin}}, \bibinfo {author} {\bibfnamefont {Y.}~\bibnamefont
			{Wang}}, \bibinfo {author} {\bibfnamefont {X.}~\bibnamefont {Hu}}, \bibinfo
		{author} {\bibfnamefont {L.}~\bibnamefont {Wang}}, \bibinfo {author}
		{\bibfnamefont {E.}~\bibnamefont {Liu}}, \bibinfo {author} {\bibfnamefont
			{G.}~\bibnamefont {Xu}}, \bibinfo {author} {\bibfnamefont {L.}~\bibnamefont
			{Lu}}, \bibinfo {author} {\bibfnamefont {A.~A.}\ \bibnamefont {Soluyanov}},
		\bibinfo {author} {\bibfnamefont {H.}~\bibnamefont {Lei}}, \bibinfo {author}
		{\bibfnamefont {Y.}~\bibnamefont {Shi}}, \bibinfo {author} {\bibfnamefont
			{J.}~\bibnamefont {Luo}},\ and\ \bibinfo {author} {\bibfnamefont {Z.~G.}\
			\bibnamefont {Chen}},\ }\bibfield  {title} {\bibinfo {title} {{Electronic
				correlations and flattened band in magnetic Weyl semimetal candidate
				Co$_3$Sn$_2$S$_2$}},\ }\bibfield  {journal} {\bibinfo  {journal} {Nature
			Communications}\ }\textbf {\bibinfo {volume} {11}},\ \href
	{https://doi.org/10.1038/s41467-020-17234-0} {10.1038/s41467-020-17234-0}
	(\bibinfo {year} {2020}),\ \Eprint {https://arxiv.org/abs/1908.04561}
	{arXiv:1908.04561} \BibitemShut {NoStop}%
	\bibitem [{\citenamefont {Yin}\ \emph {et~al.}(2019)\citenamefont {Yin},
		\citenamefont {Zhang}, \citenamefont {Chang}, \citenamefont {Wang},
		\citenamefont {Tsirkin}, \citenamefont {Guguchia}, \citenamefont {Lian},
		\citenamefont {Zhou}, \citenamefont {Jiang}, \citenamefont {Belopolski},
		\citenamefont {Shumiya}, \citenamefont {Multer}, \citenamefont {Litskevich},
		\citenamefont {Cochran}, \citenamefont {Lin}, \citenamefont {Wang},
		\citenamefont {Neupert}, \citenamefont {Jia}, \citenamefont {Lei},\ and\
		\citenamefont {Hasan}}]{Yin2019}%
	\BibitemOpen
	\bibfield  {author} {\bibinfo {author} {\bibfnamefont {J.~X.}\ \bibnamefont
			{Yin}}, \bibinfo {author} {\bibfnamefont {S.~S.}\ \bibnamefont {Zhang}},
		\bibinfo {author} {\bibfnamefont {G.}~\bibnamefont {Chang}}, \bibinfo
		{author} {\bibfnamefont {Q.}~\bibnamefont {Wang}}, \bibinfo {author}
		{\bibfnamefont {S.~S.}\ \bibnamefont {Tsirkin}}, \bibinfo {author}
		{\bibfnamefont {Z.}~\bibnamefont {Guguchia}}, \bibinfo {author}
		{\bibfnamefont {B.}~\bibnamefont {Lian}}, \bibinfo {author} {\bibfnamefont
			{H.}~\bibnamefont {Zhou}}, \bibinfo {author} {\bibfnamefont {K.}~\bibnamefont
			{Jiang}}, \bibinfo {author} {\bibfnamefont {I.}~\bibnamefont {Belopolski}},
		\bibinfo {author} {\bibfnamefont {N.}~\bibnamefont {Shumiya}}, \bibinfo
		{author} {\bibfnamefont {D.}~\bibnamefont {Multer}}, \bibinfo {author}
		{\bibfnamefont {M.}~\bibnamefont {Litskevich}}, \bibinfo {author}
		{\bibfnamefont {T.~A.}\ \bibnamefont {Cochran}}, \bibinfo {author}
		{\bibfnamefont {H.}~\bibnamefont {Lin}}, \bibinfo {author} {\bibfnamefont
			{Z.}~\bibnamefont {Wang}}, \bibinfo {author} {\bibfnamefont {T.}~\bibnamefont
			{Neupert}}, \bibinfo {author} {\bibfnamefont {S.}~\bibnamefont {Jia}},
		\bibinfo {author} {\bibfnamefont {H.}~\bibnamefont {Lei}},\ and\ \bibinfo
		{author} {\bibfnamefont {M.~Z.}\ \bibnamefont {Hasan}},\ }\bibfield  {title}
	{\bibinfo {title} {{Negative flat band magnetism in a spin–orbit-coupled
				correlated kagome magnet}},\ }\href
	{https://doi.org/10.1038/s41567-019-0426-7} {\bibfield  {journal} {\bibinfo
			{journal} {Nature Physics}\ }\textbf {\bibinfo {volume} {15}},\ \bibinfo
		{pages} {443} (\bibinfo {year} {2019})}\BibitemShut {NoStop}%
	\bibitem [{\citenamefont {Mielke}\ \emph {et~al.}(2021)\citenamefont {Mielke},
		\citenamefont {Qin}, \citenamefont {Yin}, \citenamefont {Nakamura},
		\citenamefont {Das}, \citenamefont {Guo}, \citenamefont {Khasanov},
		\citenamefont {Chang}, \citenamefont {Wang}, \citenamefont {Jia},
		\citenamefont {Nakatsuji}, \citenamefont {Amato}, \citenamefont {Luetkens},
		\citenamefont {Xu}, \citenamefont {Hasan},\ and\ \citenamefont
		{Guguchia}}]{Mielke2021}%
	\BibitemOpen
	\bibfield  {author} {\bibinfo {author} {\bibfnamefont {C.}~\bibnamefont
			{Mielke}}, \bibinfo {author} {\bibfnamefont {Y.}~\bibnamefont {Qin}},
		\bibinfo {author} {\bibfnamefont {J.-X.}\ \bibnamefont {Yin}}, \bibinfo
		{author} {\bibfnamefont {H.}~\bibnamefont {Nakamura}}, \bibinfo {author}
		{\bibfnamefont {D.}~\bibnamefont {Das}}, \bibinfo {author} {\bibfnamefont
			{K.}~\bibnamefont {Guo}}, \bibinfo {author} {\bibfnamefont {R.}~\bibnamefont
			{Khasanov}}, \bibinfo {author} {\bibfnamefont {J.}~\bibnamefont {Chang}},
		\bibinfo {author} {\bibfnamefont {Z.~Q.}\ \bibnamefont {Wang}}, \bibinfo
		{author} {\bibfnamefont {S.}~\bibnamefont {Jia}}, \bibinfo {author}
		{\bibfnamefont {S.}~\bibnamefont {Nakatsuji}}, \bibinfo {author}
		{\bibfnamefont {A.}~\bibnamefont {Amato}}, \bibinfo {author} {\bibfnamefont
			{H.}~\bibnamefont {Luetkens}}, \bibinfo {author} {\bibfnamefont
			{G.}~\bibnamefont {Xu}}, \bibinfo {author} {\bibfnamefont {M.~Z.}\
			\bibnamefont {Hasan}},\ and\ \bibinfo {author} {\bibfnamefont
			{Z.}~\bibnamefont {Guguchia}},\ }\bibfield  {title} {\bibinfo {title} {{
				Nodeless kagome superconductivity in LaRu$_3$Si$_2$ }},\ }\href
	{https://doi.org/10.1103/physrevmaterials.5.034803} {\bibfield  {journal}
		{\bibinfo  {journal} {Physical Review Materials}\ }\textbf {\bibinfo {volume}
			{5}},\ \bibinfo {pages} {034803} (\bibinfo {year} {2021})}\BibitemShut {NoStop}%
	\bibitem [{\citenamefont {Kang}\ \emph {et~al.}(2020)\citenamefont {Kang},
		\citenamefont {Ye}, \citenamefont {Fang}, \citenamefont {You}, \citenamefont
		{Levitan}, \citenamefont {Han}, \citenamefont {Facio}, \citenamefont
		{Jozwiak}, \citenamefont {Bostwick}, \citenamefont {Rotenberg}, \citenamefont
		{Chan}, \citenamefont {McDonald}, \citenamefont {Graf}, \citenamefont
		{Kaznatcheev}, \citenamefont {Vescovo}, \citenamefont {Bell}, \citenamefont
		{Kaxiras}, \citenamefont {van~den Brink}, \citenamefont {Richter},
		\citenamefont {{Prasad Ghimire}}, \citenamefont {Checkelsky},\ and\
		\citenamefont {Comin}}]{Kang2020}%
	\BibitemOpen
	\bibfield  {author} {\bibinfo {author} {\bibfnamefont {M.}~\bibnamefont
			{Kang}}, \bibinfo {author} {\bibfnamefont {L.}~\bibnamefont {Ye}}, \bibinfo
		{author} {\bibfnamefont {S.}~\bibnamefont {Fang}}, \bibinfo {author}
		{\bibfnamefont {J.~S.}\ \bibnamefont {You}}, \bibinfo {author} {\bibfnamefont
			{A.}~\bibnamefont {Levitan}}, \bibinfo {author} {\bibfnamefont
			{M.}~\bibnamefont {Han}}, \bibinfo {author} {\bibfnamefont {J.~I.}\
			\bibnamefont {Facio}}, \bibinfo {author} {\bibfnamefont {C.}~\bibnamefont
			{Jozwiak}}, \bibinfo {author} {\bibfnamefont {A.}~\bibnamefont {Bostwick}},
		\bibinfo {author} {\bibfnamefont {E.}~\bibnamefont {Rotenberg}}, \bibinfo
		{author} {\bibfnamefont {M.~K.}\ \bibnamefont {Chan}}, \bibinfo {author}
		{\bibfnamefont {R.~D.}\ \bibnamefont {McDonald}}, \bibinfo {author}
		{\bibfnamefont {D.}~\bibnamefont {Graf}}, \bibinfo {author} {\bibfnamefont
			{K.}~\bibnamefont {Kaznatcheev}}, \bibinfo {author} {\bibfnamefont
			{E.}~\bibnamefont {Vescovo}}, \bibinfo {author} {\bibfnamefont {D.~C.}\
			\bibnamefont {Bell}}, \bibinfo {author} {\bibfnamefont {E.}~\bibnamefont
			{Kaxiras}}, \bibinfo {author} {\bibfnamefont {J.}~\bibnamefont {van~den
				Brink}}, \bibinfo {author} {\bibfnamefont {M.}~\bibnamefont {Richter}},
		\bibinfo {author} {\bibfnamefont {M.}~\bibnamefont {{Prasad Ghimire}}},
		\bibinfo {author} {\bibfnamefont {J.~G.}\ \bibnamefont {Checkelsky}},\ and\
		\bibinfo {author} {\bibfnamefont {R.}~\bibnamefont {Comin}},\ }\bibfield
	{title} {\bibinfo {title} {{Dirac fermions and flat bands in the ideal kagome
				metal FeSn}},\ }\href {https://doi.org/10.1038/s41563-019-0531-0} {\bibfield
		{journal} {\bibinfo  {journal} {Nature Materials}\ }\textbf {\bibinfo
			{volume} {19}},\ \bibinfo {pages} {163} (\bibinfo {year} {2020})},\ \Eprint
	{https://arxiv.org/abs/1906.02167} {arXiv:1906.02167} \BibitemShut {NoStop}%
	\bibitem [{\citenamefont {Ye}\ \emph {et~al.}(2018)\citenamefont {Ye},
		\citenamefont {Kang}, \citenamefont {Liu}, \citenamefont {von Cube},
		\citenamefont {Wicker}, \citenamefont {Suzuki}, \citenamefont {Jozwiak},
		\citenamefont {Bostwick}, \citenamefont {Rotenberg}, \citenamefont {Bell},
		\citenamefont {Fu}, \citenamefont {Comin},\ and\ \citenamefont
		{Checkelsky}}]{Ye2018}%
	\BibitemOpen
	\bibfield  {author} {\bibinfo {author} {\bibfnamefont {L.}~\bibnamefont
			{Ye}}, \bibinfo {author} {\bibfnamefont {M.}~\bibnamefont {Kang}}, \bibinfo
		{author} {\bibfnamefont {J.}~\bibnamefont {Liu}}, \bibinfo {author}
		{\bibfnamefont {F.}~\bibnamefont {von Cube}}, \bibinfo {author}
		{\bibfnamefont {C.~R.}\ \bibnamefont {Wicker}}, \bibinfo {author}
		{\bibfnamefont {T.}~\bibnamefont {Suzuki}}, \bibinfo {author} {\bibfnamefont
			{C.}~\bibnamefont {Jozwiak}}, \bibinfo {author} {\bibfnamefont
			{A.}~\bibnamefont {Bostwick}}, \bibinfo {author} {\bibfnamefont
			{E.}~\bibnamefont {Rotenberg}}, \bibinfo {author} {\bibfnamefont {D.~C.}\
			\bibnamefont {Bell}}, \bibinfo {author} {\bibfnamefont {L.}~\bibnamefont
			{Fu}}, \bibinfo {author} {\bibfnamefont {R.}~\bibnamefont {Comin}},\ and\
		\bibinfo {author} {\bibfnamefont {J.~G.}\ \bibnamefont {Checkelsky}},\
	}\bibfield  {title} {\bibinfo {title} {{Massive Dirac fermions in a
				ferromagnetic kagome metal}},\ }\href {https://doi.org/10.1038/nature25987}
	{\bibfield  {journal} {\bibinfo  {journal} {Nature}\ }\textbf {\bibinfo
			{volume} {555}},\ \bibinfo {pages} {638} (\bibinfo {year}
		{2018})}\BibitemShut {NoStop}%
	\bibitem [{\citenamefont {Armitage}\ \emph {et~al.}(2018)\citenamefont
		{Armitage}, \citenamefont {Mele},\ and\ \citenamefont
		{Vishwanath}}]{Armitage2018}%
	\BibitemOpen
	\bibfield  {author} {\bibinfo {author} {\bibfnamefont {N.~P.}\ \bibnamefont
			{Armitage}}, \bibinfo {author} {\bibfnamefont {E.~J.}\ \bibnamefont {Mele}},\
		and\ \bibinfo {author} {\bibfnamefont {A.}~\bibnamefont {Vishwanath}},\
	}\bibfield  {title} {\bibinfo {title} {{Weyl and Dirac semimetals in
				three-dimensional solids}},\ }\href
	{https://doi.org/10.1103/RevModPhys.90.015001} {\bibfield  {journal}
		{\bibinfo  {journal} {Reviews of Modern Physics}\ }\textbf {\bibinfo {volume}
			{90}},\ \bibinfo {pages} {15001} (\bibinfo {year} {2018})}\BibitemShut
	{NoStop}%
	\bibitem [{\citenamefont {Dedkov}\ \emph {et~al.}(2008)\citenamefont {Dedkov},
		\citenamefont {Holder}, \citenamefont {Molodtsov},\ and\ \citenamefont
		{Rosner}}]{Dedkov2008}%
	\BibitemOpen
	\bibfield  {author} {\bibinfo {author} {\bibfnamefont {Y.~S.}\ \bibnamefont
			{Dedkov}}, \bibinfo {author} {\bibfnamefont {M.}~\bibnamefont {Holder}},
		\bibinfo {author} {\bibfnamefont {S.~L.}\ \bibnamefont {Molodtsov}},\ and\
		\bibinfo {author} {\bibfnamefont {H.}~\bibnamefont {Rosner}},\ }\bibfield
	{title} {\bibinfo {title} {{Electronic structure of shandite
				Co$_3$Sn$_2$S$_2$}},\ }\bibfield  {journal} {\bibinfo  {journal} {Journal of
			Physics: Conference Series}\ }\textbf {\bibinfo {volume} {100}},\ \href
	{https://doi.org/10.1088/1742-6596/100/7/072011}
	{10.1088/1742-6596/100/7/072011} (\bibinfo {year} {2008})\BibitemShut
	{NoStop}%
	\bibitem [{\citenamefont {Zhang}\ \emph {et~al.}(2020)\citenamefont {Zhang},
		\citenamefont {Yin}, \citenamefont {Ikhlas}, \citenamefont {Tien},
		\citenamefont {Wang}, \citenamefont {Shumiya}, \citenamefont {Chang},
		\citenamefont {Tsirkin}, \citenamefont {Shi}, \citenamefont {Yi},
		\citenamefont {Guguchia}, \citenamefont {Li}, \citenamefont {Wang},
		\citenamefont {Chang}, \citenamefont {Wang}, \citenamefont {Yang},
		\citenamefont {Neupert}, \citenamefont {Nakatsuji},\ and\ \citenamefont
		{Hasan}}]{Zhang2020}%
	\BibitemOpen
	\bibfield  {author} {\bibinfo {author} {\bibfnamefont {S.~S.}\ \bibnamefont
			{Zhang}}, \bibinfo {author} {\bibfnamefont {J.~X.}\ \bibnamefont {Yin}},
		\bibinfo {author} {\bibfnamefont {M.}~\bibnamefont {Ikhlas}}, \bibinfo
		{author} {\bibfnamefont {H.~J.}\ \bibnamefont {Tien}}, \bibinfo {author}
		{\bibfnamefont {R.}~\bibnamefont {Wang}}, \bibinfo {author} {\bibfnamefont
			{N.}~\bibnamefont {Shumiya}}, \bibinfo {author} {\bibfnamefont
			{G.}~\bibnamefont {Chang}}, \bibinfo {author} {\bibfnamefont {S.~S.}\
			\bibnamefont {Tsirkin}}, \bibinfo {author} {\bibfnamefont {Y.}~\bibnamefont
			{Shi}}, \bibinfo {author} {\bibfnamefont {C.}~\bibnamefont {Yi}}, \bibinfo
		{author} {\bibfnamefont {Z.}~\bibnamefont {Guguchia}}, \bibinfo {author}
		{\bibfnamefont {H.}~\bibnamefont {Li}}, \bibinfo {author} {\bibfnamefont
			{W.}~\bibnamefont {Wang}}, \bibinfo {author} {\bibfnamefont {T.~R.}\
			\bibnamefont {Chang}}, \bibinfo {author} {\bibfnamefont {Z.}~\bibnamefont
			{Wang}}, \bibinfo {author} {\bibfnamefont {Y.~F.}\ \bibnamefont {Yang}},
		\bibinfo {author} {\bibfnamefont {T.}~\bibnamefont {Neupert}}, \bibinfo
		{author} {\bibfnamefont {S.}~\bibnamefont {Nakatsuji}},\ and\ \bibinfo
		{author} {\bibfnamefont {M.~Z.}\ \bibnamefont {Hasan}},\ }\bibfield  {title}
	{\bibinfo {title} {{Many-Body Resonance in a Correlated Topological Kagome
				Antiferromagnet}},\ }\href {https://doi.org/10.1103/PhysRevLett.125.046401}
	{\bibfield  {journal} {\bibinfo  {journal} {Physical Review Letters}\
		}\textbf {\bibinfo {volume} {125}},\ \bibinfo {pages} {046401} (\bibinfo
		{year} {2020})}\BibitemShut {NoStop}%
	\bibitem [{\citenamefont {Hou}\ \emph {et~al.}(2017)\citenamefont {Hou},
		\citenamefont {Ren}, \citenamefont {Ding}, \citenamefont {Xu}, \citenamefont
		{Wang}, \citenamefont {Yang}, \citenamefont {Zhang}, \citenamefont {Zhang},
		\citenamefont {Liu}, \citenamefont {Xu}, \citenamefont {Wang}, \citenamefont
		{Wu}, \citenamefont {Zhang}, \citenamefont {Shen},\ and\ \citenamefont
		{Zhang}}]{Hou2017}%
	\BibitemOpen
	\bibfield  {author} {\bibinfo {author} {\bibfnamefont {Z.}~\bibnamefont
			{Hou}}, \bibinfo {author} {\bibfnamefont {W.}~\bibnamefont {Ren}}, \bibinfo
		{author} {\bibfnamefont {B.}~\bibnamefont {Ding}}, \bibinfo {author}
		{\bibfnamefont {G.}~\bibnamefont {Xu}}, \bibinfo {author} {\bibfnamefont
			{Y.}~\bibnamefont {Wang}}, \bibinfo {author} {\bibfnamefont {B.}~\bibnamefont
			{Yang}}, \bibinfo {author} {\bibfnamefont {Q.}~\bibnamefont {Zhang}},
		\bibinfo {author} {\bibfnamefont {Y.}~\bibnamefont {Zhang}}, \bibinfo
		{author} {\bibfnamefont {E.}~\bibnamefont {Liu}}, \bibinfo {author}
		{\bibfnamefont {F.}~\bibnamefont {Xu}}, \bibinfo {author} {\bibfnamefont
			{W.}~\bibnamefont {Wang}}, \bibinfo {author} {\bibfnamefont {G.}~\bibnamefont
			{Wu}}, \bibinfo {author} {\bibfnamefont {X.}~\bibnamefont {Zhang}}, \bibinfo
		{author} {\bibfnamefont {B.}~\bibnamefont {Shen}},\ and\ \bibinfo {author}
		{\bibfnamefont {Z.}~\bibnamefont {Zhang}},\ }\bibfield  {title} {\bibinfo
		{title} {{Observation of Various and Spontaneous Magnetic Skyrmionic Bubbles
				at Room Temperature in a Frustrated Kagome Magnet with Uniaxial Magnetic
				Anisotropy}},\ }\href {https://doi.org/10.1002/adma.201701144} {\bibfield
		{journal} {\bibinfo  {journal} {Advanced Materials}\ }\textbf {\bibinfo
			{volume} {29}},\ \bibinfo {pages} {1701144} (\bibinfo {year}
		{2017})}\BibitemShut {NoStop}%
	\bibitem [{\citenamefont {Pereiro}\ \emph {et~al.}(2014)\citenamefont
		{Pereiro}, \citenamefont {Yudin}, \citenamefont {Chico}, \citenamefont {Etz},
		\citenamefont {Eriksson},\ and\ \citenamefont {Bergman}}]{Pereiro2014}%
	\BibitemOpen
	\bibfield  {author} {\bibinfo {author} {\bibfnamefont {M.}~\bibnamefont
			{Pereiro}}, \bibinfo {author} {\bibfnamefont {D.}~\bibnamefont {Yudin}},
		\bibinfo {author} {\bibfnamefont {J.}~\bibnamefont {Chico}}, \bibinfo
		{author} {\bibfnamefont {C.}~\bibnamefont {Etz}}, \bibinfo {author}
		{\bibfnamefont {O.}~\bibnamefont {Eriksson}},\ and\ \bibinfo {author}
		{\bibfnamefont {A.}~\bibnamefont {Bergman}},\ }\bibfield  {title} {\bibinfo
		{title} {{Topological excitations in a kagome magnet}},\ }\bibfield
	{journal} {\bibinfo  {journal} {Nature Communications}\ }\textbf {\bibinfo
		{volume} {5}},\ \href {https://doi.org/10.1038/ncomms5815}
	{10.1038/ncomms5815} (\bibinfo {year} {2014}),\ \Eprint
	{https://arxiv.org/abs/1401.4757} {arXiv:1401.4757} \BibitemShut {NoStop}%
	\bibitem [{\citenamefont {Yin}\ \emph {et~al.}(2021)\citenamefont {Yin},
		\citenamefont {Tu}, \citenamefont {Gong}, \citenamefont {Fu}, \citenamefont
		{Yan},\ and\ \citenamefont {Lei}}]{Yin2021}%
	\BibitemOpen
	\bibfield  {author} {\bibinfo {author} {\bibfnamefont {Q.}~\bibnamefont
			{Yin}}, \bibinfo {author} {\bibfnamefont {Z.}~\bibnamefont {Tu}}, \bibinfo
		{author} {\bibfnamefont {C.}~\bibnamefont {Gong}}, \bibinfo {author}
		{\bibfnamefont {Y.}~\bibnamefont {Fu}}, \bibinfo {author} {\bibfnamefont
			{S.}~\bibnamefont {Yan}},\ and\ \bibinfo {author} {\bibfnamefont
			{H.}~\bibnamefont {Lei}},\ }\bibfield  {title} {\bibinfo {title}
		{{Superconductivity and Normal-State Properties of Kagome Metal RbV$_3$Sb$_5$
				Single Crystals}},\ }\bibfield  {journal} {\bibinfo  {journal} {Chinese
			Physics Letters}\ }\textbf {\bibinfo {volume} {38}},\ \href
	{https://doi.org/10.1088/0256-307X/38/3/037403}
	{10.1088/0256-307X/38/3/037403} (\bibinfo {year} {2021}),\ \Eprint
	{https://arxiv.org/abs/2101.10193} {arXiv:2101.10193} \BibitemShut {NoStop}%
	\bibitem [{\citenamefont {Ortiz}\ \emph {et~al.}(2021)\citenamefont {Ortiz},
		\citenamefont {Sarte}, \citenamefont {Kenney}, \citenamefont {Graf},
		\citenamefont {Teicher}, \citenamefont {Seshadri},\ and\ \citenamefont
		{Wilson}}]{Ortiz2021}%
	\BibitemOpen
	\bibfield  {author} {\bibinfo {author} {\bibfnamefont {B.~R.}\ \bibnamefont
			{Ortiz}}, \bibinfo {author} {\bibfnamefont {P.~M.}\ \bibnamefont {Sarte}},
		\bibinfo {author} {\bibfnamefont {E.~M.}\ \bibnamefont {Kenney}}, \bibinfo
		{author} {\bibfnamefont {M.~J.}\ \bibnamefont {Graf}}, \bibinfo {author}
		{\bibfnamefont {S.~M.~L.}\ \bibnamefont {Teicher}}, \bibinfo {author}
		{\bibfnamefont {R.}~\bibnamefont {Seshadri}},\ and\ \bibinfo {author}
		{\bibfnamefont {S.~D.}\ \bibnamefont {Wilson}},\ }\bibfield  {title}
	{\bibinfo {title} {{ Superconductivity in the Z$_2$ kagomé metal
				KV$_3$Sb$_5$}},\ }\href {https://doi.org/10.1103/physrevmaterials.5.034801}
	{\bibfield  {journal} {\bibinfo  {journal} {Physical Review Materials}\
		}\textbf {\bibinfo {volume} {5}},\ \bibinfo {pages} {034801} (\bibinfo {year}
		{2021})},\ \Eprint {https://arxiv.org/abs/arXiv:2012.09097v2}
	{arXiv:arXiv:2012.09097v2} \BibitemShut {NoStop}%
	\bibitem [{\citenamefont {Ortiz}\ \emph {et~al.}(2020)\citenamefont {Ortiz},
		\citenamefont {Teicher}, \citenamefont {Hu}, \citenamefont {Zuo},
		\citenamefont {Sarte}, \citenamefont {Schueller}, \citenamefont {Abeykoon},
		\citenamefont {Krogstad}, \citenamefont {Rosenkranz}, \citenamefont {Osborn},
		\citenamefont {Seshadri}, \citenamefont {Balents}, \citenamefont {He},\ and\
		\citenamefont {Wilson}}]{Ortiz2020}%
	\BibitemOpen
	\bibfield  {author} {\bibinfo {author} {\bibfnamefont {B.~R.}\ \bibnamefont
			{Ortiz}}, \bibinfo {author} {\bibfnamefont {S.~M.~L.}\ \bibnamefont {Teicher}},
		\bibinfo {author} {\bibfnamefont {Y.}~\bibnamefont {Hu}}, \bibinfo {author}
		{\bibfnamefont {J.~L.}\ \bibnamefont {Zuo}}, \bibinfo {author} {\bibfnamefont
			{P.~M.}\ \bibnamefont {Sarte}}, \bibinfo {author} {\bibfnamefont {E.~C.}\
			\bibnamefont {Schueller}}, \bibinfo {author} {\bibfnamefont {A.~M.~M.}\
			\bibnamefont {Abeykoon}}, \bibinfo {author} {\bibfnamefont {M.~J.}\
			\bibnamefont {Krogstad}}, \bibinfo {author} {\bibfnamefont {S.}~\bibnamefont
			{Rosenkranz}}, \bibinfo {author} {\bibfnamefont {R.}~\bibnamefont {Osborn}},
		\bibinfo {author} {\bibfnamefont {R.}~\bibnamefont {Seshadri}}, \bibinfo
		{author} {\bibfnamefont {L.}~\bibnamefont {Balents}}, \bibinfo {author}
		{\bibfnamefont {J.}~\bibnamefont {He}},\ and\ \bibinfo {author}
		{\bibfnamefont {S.~D.}\ \bibnamefont {Wilson}},\ }\bibfield  {title}
	{\bibinfo {title} {{CsV$_3$Sb$_5$: A Z$_2$ Topological Kagomé Metal with a
				Superconducting Ground State}},\ }\href
	{https://doi.org/10.1103/PhysRevLett.125.247002} {\bibfield  {journal}
		{\bibinfo  {journal} {Physical Review Letters}\ }\textbf {\bibinfo {volume}
			{125}},\ \bibinfo {pages} {247002} (\bibinfo {year} {2020})}\BibitemShut
	{NoStop}%
	\bibitem [{\citenamefont {Uykur}\ \emph {et~al.}(2021)\citenamefont {Uykur},
		\citenamefont {Ortiz}, \citenamefont {Iakutkina}, \citenamefont {Wenzel},
		\citenamefont {Wilson}, \citenamefont {Dressel},\ and\ \citenamefont
		{Tsirlin}}]{Uykur2021}%
	\BibitemOpen
	\bibfield  {author} {\bibinfo {author} {\bibfnamefont {E.}~\bibnamefont
			{Uykur}}, \bibinfo {author} {\bibfnamefont {B.~R.}\ \bibnamefont {Ortiz}},
		\bibinfo {author} {\bibfnamefont {O.}~\bibnamefont {Iakutkina}}, \bibinfo
		{author} {\bibfnamefont {M.}~\bibnamefont {Wenzel}}, \bibinfo {author}
		{\bibfnamefont {S.~D.}\ \bibnamefont {Wilson}}, \bibinfo {author}
		{\bibfnamefont {M.}~\bibnamefont {Dressel}},\ and\ \bibinfo {author}
		{\bibfnamefont {A.~A.}\ \bibnamefont {Tsirlin}},\ }\bibfield  {title}
	{\bibinfo {title} {{Low-energy optical properties of the nonmagnetic kagome
				metal CsV$_3$Sb$_5$}},\ }\href {https://doi.org/10.1103/PhysRevB.104.045130}
	{\bibfield  {journal} {\bibinfo  {journal} {Physical Review B}\ }\textbf
		{\bibinfo {volume} {104}},\ \bibinfo {pages} {045130} (\bibinfo {year}
		{2021})}\BibitemShut {NoStop}%
	\bibitem [{\citenamefont {Uykur}\ \emph {et~al.}(2022)\citenamefont {Uykur},
		\citenamefont {Ortiz}, \citenamefont {Wilson}, \citenamefont {Dressel},\ and\
		\citenamefont {Tsirlin}}]{Uykur2022}%
	\BibitemOpen
	\bibfield  {author} {\bibinfo {author} {\bibfnamefont {E.}~\bibnamefont
			{Uykur}}, \bibinfo {author} {\bibfnamefont {B.~R.}\ \bibnamefont {Ortiz}},
		\bibinfo {author} {\bibfnamefont {S.~D.}\ \bibnamefont {Wilson}}, \bibinfo
		{author} {\bibfnamefont {M.}~\bibnamefont {Dressel}},\ and\ \bibinfo {author}
		{\bibfnamefont {A.~A.}\ \bibnamefont {Tsirlin}},\ }\bibfield  {title}
	{\bibinfo {title} {{Optical detection of the density-wave instability in the
				kagome metal KV$_3$Sb$_5$}},\ }\href
	{https://doi.org/10.1038/s41535-021-00420-8} {\bibfield  {journal} {\bibinfo
			{journal} {npj Quantum Materials}\ }\textbf {\bibinfo {volume} {7}},\
		\bibinfo {pages} {1} (\bibinfo {year} {2022})}\BibitemShut {NoStop}%
	\bibitem [{\citenamefont {Nagaosa}\ \emph {et~al.}(2010)\citenamefont
		{Nagaosa}, \citenamefont {Sinova}, \citenamefont {Onoda}, \citenamefont
		{MacDonald},\ and\ \citenamefont {Ong}}]{Nagaosa2010}%
	\BibitemOpen
	\bibfield  {author} {\bibinfo {author} {\bibfnamefont {N.}~\bibnamefont
			{Nagaosa}}, \bibinfo {author} {\bibfnamefont {J.}~\bibnamefont {Sinova}},
		\bibinfo {author} {\bibfnamefont {S.}~\bibnamefont {Onoda}}, \bibinfo
		{author} {\bibfnamefont {A.~H.}\ \bibnamefont {MacDonald}},\ and\ \bibinfo
		{author} {\bibfnamefont {N.~P.}\ \bibnamefont {Ong}},\ }\bibfield  {title}
	{\bibinfo {title} {{Anomalous Hall effect}},\ }\href
	{https://doi.org/10.1103/RevModPhys.82.1539} {\bibfield  {journal} {\bibinfo
			{journal} {Reviews of Modern Physics}\ }\textbf {\bibinfo {volume} {82}},\
		\bibinfo {pages} {1539} (\bibinfo {year} {2010})},\ \Eprint
	{https://arxiv.org/abs/arXiv:0904.4154v1} {arXiv:arXiv:0904.4154v1}
	\BibitemShut {NoStop}%
	\bibitem [{\citenamefont {Fang}\ \emph {et~al.}(2003)\citenamefont {Fang},
		\citenamefont {Nagaosa}, \citenamefont {Takahashi}, \citenamefont {Asamitsu},
		\citenamefont {Mathieu}, \citenamefont {Ogasawara}, \citenamefont {Yamada},
		\citenamefont {Kawasaki}, \citenamefont {Tokura},\ and\ \citenamefont
		{Terakura}}]{Fang2003}%
	\BibitemOpen
	\bibfield  {author} {\bibinfo {author} {\bibfnamefont {Z.}~\bibnamefont
			{Fang}}, \bibinfo {author} {\bibfnamefont {N.}~\bibnamefont {Nagaosa}},
		\bibinfo {author} {\bibfnamefont {K.~S.}\ \bibnamefont {Takahashi}}, \bibinfo
		{author} {\bibfnamefont {A.}~\bibnamefont {Asamitsu}}, \bibinfo {author}
		{\bibfnamefont {R.}~\bibnamefont {Mathieu}}, \bibinfo {author} {\bibfnamefont
			{T.}~\bibnamefont {Ogasawara}}, \bibinfo {author} {\bibfnamefont
			{H.}~\bibnamefont {Yamada}}, \bibinfo {author} {\bibfnamefont
			{M.}~\bibnamefont {Kawasaki}}, \bibinfo {author} {\bibfnamefont
			{Y.}~\bibnamefont {Tokura}},\ and\ \bibinfo {author} {\bibfnamefont
			{K.}~\bibnamefont {Terakura}},\ }\bibfield  {title} {\bibinfo {title} {The
			anomalous hall effect and magnetic monopoles in momentum space},\ }\href
	{https://doi.org/10.1126/science.1089408} {\bibfield  {journal} {\bibinfo
			{journal} {Science}\ }\textbf {\bibinfo {volume} {302}},\ \bibinfo {pages}
		{92} (\bibinfo {year} {2003})},\ \Eprint
	{https://arxiv.org/abs/https://science.sciencemag.org/content/302/5642/92.full.pdf}
	{https://science.sciencemag.org/content/302/5642/92.full.pdf} \BibitemShut
	{NoStop}%
	\bibitem [{\citenamefont {Liu}\ \emph {et~al.}(2018)\citenamefont {Liu},
		\citenamefont {Sun}, \citenamefont {Kumar}, \citenamefont {Muechler},
		\citenamefont {Sun}, \citenamefont {Jiao}, \citenamefont {Yang},
		\citenamefont {Liu}, \citenamefont {Liang}, \citenamefont {Xu}, \citenamefont
		{Kroder}, \citenamefont {S{\"{u}}{\ss}}, \citenamefont {Borrmann},
		\citenamefont {Shekhar}, \citenamefont {Wang}, \citenamefont {Xi},
		\citenamefont {Wang}, \citenamefont {Schnelle}, \citenamefont {Wirth},
		\citenamefont {Chen}, \citenamefont {Goennenwein},\ and\ \citenamefont
		{Felser}}]{Liu2018}%
	\BibitemOpen
	\bibfield  {author} {\bibinfo {author} {\bibfnamefont {E.}~\bibnamefont
			{Liu}}, \bibinfo {author} {\bibfnamefont {Y.}~\bibnamefont {Sun}}, \bibinfo
		{author} {\bibfnamefont {N.}~\bibnamefont {Kumar}}, \bibinfo {author}
		{\bibfnamefont {L.}~\bibnamefont {Muechler}}, \bibinfo {author}
		{\bibfnamefont {A.}~\bibnamefont {Sun}}, \bibinfo {author} {\bibfnamefont
			{L.}~\bibnamefont {Jiao}}, \bibinfo {author} {\bibfnamefont {S.~Y.}\
			\bibnamefont {Yang}}, \bibinfo {author} {\bibfnamefont {D.}~\bibnamefont
			{Liu}}, \bibinfo {author} {\bibfnamefont {A.}~\bibnamefont {Liang}}, \bibinfo
		{author} {\bibfnamefont {Q.}~\bibnamefont {Xu}}, \bibinfo {author}
		{\bibfnamefont {J.}~\bibnamefont {Kroder}}, \bibinfo {author} {\bibfnamefont
			{V.}~\bibnamefont {S{\"{u}}{\ss}}}, \bibinfo {author} {\bibfnamefont
			{H.}~\bibnamefont {Borrmann}}, \bibinfo {author} {\bibfnamefont
			{C.}~\bibnamefont {Shekhar}}, \bibinfo {author} {\bibfnamefont
			{Z.}~\bibnamefont {Wang}}, \bibinfo {author} {\bibfnamefont {C.}~\bibnamefont
			{Xi}}, \bibinfo {author} {\bibfnamefont {W.}~\bibnamefont {Wang}}, \bibinfo
		{author} {\bibfnamefont {W.}~\bibnamefont {Schnelle}}, \bibinfo {author}
		{\bibfnamefont {S.}~\bibnamefont {Wirth}}, \bibinfo {author} {\bibfnamefont
			{Y.}~\bibnamefont {Chen}}, \bibinfo {author} {\bibfnamefont {S.~T.}\
			\bibnamefont {Goennenwein}},\ and\ \bibinfo {author} {\bibfnamefont
			{C.}~\bibnamefont {Felser}},\ }\bibfield  {title} {\bibinfo {title} {{Giant
				anomalous Hall effect in a ferromagnetic kagome-lattice semimetal}},\ }\href
	{https://doi.org/10.1038/s41567-018-0234-5} {\bibfield  {journal} {\bibinfo
			{journal} {Nature Physics}\ }\textbf {\bibinfo {volume} {14}},\ \bibinfo
		{pages} {1125} (\bibinfo {year} {2018})}\BibitemShut {NoStop}%
	\bibitem [{\citenamefont {Burkov}\ and\ \citenamefont
		{Balents}(2011)}]{Burkov2011}%
	\BibitemOpen
	\bibfield  {author} {\bibinfo {author} {\bibfnamefont {A.~A.}\ \bibnamefont
			{Burkov}}\ and\ \bibinfo {author} {\bibfnamefont {L.}~\bibnamefont
			{Balents}},\ }\bibfield  {title} {\bibinfo {title} {{Weyl semimetal in a
				topological insulator multilayer}},\ }\href
	{https://doi.org/10.1103/PhysRevLett.107.127205} {\bibfield  {journal}
		{\bibinfo  {journal} {Physical Review Letters}\ }\textbf {\bibinfo {volume}
			{107}},\ \bibinfo {pages} {127205} (\bibinfo {year} {2011})},\ \Eprint
	{https://arxiv.org/abs/1105.5138} {arXiv:1105.5138} \BibitemShut {NoStop}%
	\bibitem [{\citenamefont {Wang}\ \emph {et~al.}(2016)\citenamefont {Wang},
		\citenamefont {Sun}, \citenamefont {Zhang}, \citenamefont {Pang},\ and\
		\citenamefont {Lei}}]{Wang2016}%
	\BibitemOpen
	\bibfield  {author} {\bibinfo {author} {\bibfnamefont {Q.}~\bibnamefont
			{Wang}}, \bibinfo {author} {\bibfnamefont {S.}~\bibnamefont {Sun}}, \bibinfo
		{author} {\bibfnamefont {X.}~\bibnamefont {Zhang}}, \bibinfo {author}
		{\bibfnamefont {F.}~\bibnamefont {Pang}},\ and\ \bibinfo {author}
		{\bibfnamefont {H.}~\bibnamefont {Lei}},\ }\bibfield  {title} {\bibinfo
		{title} {{Anomalous Hall effect in a ferromagnetic Fe$_3$Sn$_2$ single
				crystal with a geometrically frustrated Fe bilayer kagome lattice}},\ }\href
	{https://doi.org/10.1103/PhysRevB.94.075135} {\bibfield  {journal} {\bibinfo
			{journal} {Physical Review B}\ }\textbf {\bibinfo {volume} {94}},\ \bibinfo
		{pages} {075135} (\bibinfo {year} {2016})},\ \Eprint
	{https://arxiv.org/abs/arXiv:1610.04970v1} {arXiv:arXiv:1610.04970v1}
	\BibitemShut {NoStop}%
	\bibitem [{\citenamefont {Biswas}\ \emph {et~al.}(2020)\citenamefont {Biswas},
		\citenamefont {Iakutkina}, \citenamefont {Wang}, \citenamefont {Lei},
		\citenamefont {Dressel},\ and\ \citenamefont {Uykur}}]{Biswas2020}%
	\BibitemOpen
	\bibfield  {author} {\bibinfo {author} {\bibfnamefont {A.}~\bibnamefont
			{Biswas}}, \bibinfo {author} {\bibfnamefont {O.}~\bibnamefont {Iakutkina}},
		\bibinfo {author} {\bibfnamefont {Q.}~\bibnamefont {Wang}}, \bibinfo {author}
		{\bibfnamefont {H.~C.}\ \bibnamefont {Lei}}, \bibinfo {author} {\bibfnamefont
			{M.}~\bibnamefont {Dressel}},\ and\ \bibinfo {author} {\bibfnamefont
			{E.}~\bibnamefont {Uykur}},\ }\bibfield  {title} {\bibinfo {title}
		{{Spin-Reorientation-Induced Band Gap in Fe$_3$Sn$_2$: Optical Signatures of
				Weyl nodes}},\ }\href {https://doi.org/10.1103/PhysRevLett.125.076403}
	{\bibfield  {journal} {\bibinfo  {journal} {Physical Review Letters}\
		}\textbf {\bibinfo {volume} {125}},\ \bibinfo {pages} {076403} (\bibinfo
		{year} {2020})},\ \Eprint {https://arxiv.org/abs/2007.14828}
	{arXiv:2007.14828} \BibitemShut {NoStop}%
	\bibitem [{\citenamefont {Yin}\ \emph {et~al.}(2018)\citenamefont {Yin},
		\citenamefont {Zhang}, \citenamefont {Li}, \citenamefont {Jiang},
		\citenamefont {Chang}, \citenamefont {Zhang}, \citenamefont {Lian},
		\citenamefont {Xiang}, \citenamefont {Belopolski}, \citenamefont {Zheng},
		\citenamefont {Cochran}, \citenamefont {Xu}, \citenamefont {Bian},
		\citenamefont {Liu}, \citenamefont {Chang}, \citenamefont {Lin},
		\citenamefont {Lu}, \citenamefont {Wang}, \citenamefont {Jia}, \citenamefont
		{Wang},\ and\ \citenamefont {Hasan}}]{Yin2018}%
	\BibitemOpen
	\bibfield  {author} {\bibinfo {author} {\bibfnamefont {J.~X.}\ \bibnamefont
			{Yin}}, \bibinfo {author} {\bibfnamefont {S.~S.}\ \bibnamefont {Zhang}},
		\bibinfo {author} {\bibfnamefont {H.}~\bibnamefont {Li}}, \bibinfo {author}
		{\bibfnamefont {K.}~\bibnamefont {Jiang}}, \bibinfo {author} {\bibfnamefont
			{G.}~\bibnamefont {Chang}}, \bibinfo {author} {\bibfnamefont
			{B.}~\bibnamefont {Zhang}}, \bibinfo {author} {\bibfnamefont
			{B.}~\bibnamefont {Lian}}, \bibinfo {author} {\bibfnamefont {C.}~\bibnamefont
			{Xiang}}, \bibinfo {author} {\bibfnamefont {I.}~\bibnamefont {Belopolski}},
		\bibinfo {author} {\bibfnamefont {H.}~\bibnamefont {Zheng}}, \bibinfo
		{author} {\bibfnamefont {T.~A.}\ \bibnamefont {Cochran}}, \bibinfo {author}
		{\bibfnamefont {S.~Y.}\ \bibnamefont {Xu}}, \bibinfo {author} {\bibfnamefont
			{G.}~\bibnamefont {Bian}}, \bibinfo {author} {\bibfnamefont {K.}~\bibnamefont
			{Liu}}, \bibinfo {author} {\bibfnamefont {T.~R.}\ \bibnamefont {Chang}},
		\bibinfo {author} {\bibfnamefont {H.}~\bibnamefont {Lin}}, \bibinfo {author}
		{\bibfnamefont {Z.~Y.}\ \bibnamefont {Lu}}, \bibinfo {author} {\bibfnamefont
			{Z.}~\bibnamefont {Wang}}, \bibinfo {author} {\bibfnamefont {S.}~\bibnamefont
			{Jia}}, \bibinfo {author} {\bibfnamefont {W.}~\bibnamefont {Wang}},\ and\
		\bibinfo {author} {\bibfnamefont {M.~Z.}\ \bibnamefont {Hasan}},\ }\bibfield
	{title} {\bibinfo {title} {{Giant and anisotropic many-body spin–orbit
				tunability in a strongly correlated kagome magnet}},\ }\href
	{https://doi.org/10.1038/s41586-018-0502-7} {\bibfield  {journal} {\bibinfo
			{journal} {Nature}\ }\textbf {\bibinfo {volume} {562}},\ \bibinfo {pages}
		{91} (\bibinfo {year} {2018})}\BibitemShut {NoStop}%
	\bibitem [{\citenamefont {Le~Caer}\ \emph {et~al.}(1978)\citenamefont
		{Le~Caer}, \citenamefont {Malaman},\ and\ \citenamefont
		{Fruchart}}]{Caer1978}%
	\BibitemOpen
	\bibfield  {author} {\bibinfo {author} {\bibfnamefont {G.}~\bibnamefont
			{Le~Caer}}, \bibinfo {author} {\bibfnamefont {B.}~\bibnamefont {Malaman}},\
		and\ \bibinfo {author} {\bibfnamefont {D.}~\bibnamefont {Fruchart}},\
	}\bibfield  {title} {\bibinfo {title} {{Magnetic properties of Fe$_3$Sn$_2$ .
				II . Neutron diffraction study (and Mossbauer effect)}},\ }\href@noop {}
	{\bibfield  {journal} {\bibinfo  {journal} {Journal of Physics F: Metal
				Physics Magnetic}\ }\textbf {\bibinfo {volume} {8}},\ \bibinfo {pages} {2389}
		(\bibinfo {year} {1978})}\BibitemShut {NoStop}%
	\bibitem [{\citenamefont {Heritage}\ \emph {et~al.}(2020)\citenamefont
		{Heritage}, \citenamefont {Bryant}, \citenamefont {Fenner}, \citenamefont
		{Wills}, \citenamefont {Aeppli},\ and\ \citenamefont {Soh}}]{Heritage2020}%
	\BibitemOpen
	\bibfield  {author} {\bibinfo {author} {\bibfnamefont {K.}~\bibnamefont
			{Heritage}}, \bibinfo {author} {\bibfnamefont {B.}~\bibnamefont {Bryant}},
		\bibinfo {author} {\bibfnamefont {L.~A.}\ \bibnamefont {Fenner}}, \bibinfo
		{author} {\bibfnamefont {A.~S.}\ \bibnamefont {Wills}}, \bibinfo {author}
		{\bibfnamefont {G.}~\bibnamefont {Aeppli}},\ and\ \bibinfo {author}
		{\bibfnamefont {Y.~A.}\ \bibnamefont {Soh}},\ }\bibfield  {title} {\bibinfo
		{title} {{Images of a First-Order Spin-Reorientation Phase Transition in a
				Metallic Kagome Ferromagnet}},\ }\href
	{https://doi.org/10.1002/adfm.201909163} {\bibfield  {journal} {\bibinfo
			{journal} {Advanced Functional Materials}\ }\textbf {\bibinfo {volume}
			{30}},\ \bibinfo {pages} {1} (\bibinfo {year} {2020})}\BibitemShut {NoStop}%
	\bibitem [{\citenamefont {Fenner}\ \emph {et~al.}(2009)\citenamefont {Fenner},
		\citenamefont {Dee},\ and\ \citenamefont {Wills}}]{Fenner2009}%
	\BibitemOpen
	\bibfield  {author} {\bibinfo {author} {\bibfnamefont {L.~A.}\ \bibnamefont
			{Fenner}}, \bibinfo {author} {\bibfnamefont {A.~A.}\ \bibnamefont {Dee}},\
		and\ \bibinfo {author} {\bibfnamefont {A.~S.}\ \bibnamefont {Wills}},\
	}\bibfield  {title} {\bibinfo {title} {{Non-collinearity and spin frustration
				in the itinerant kagome ferromagnet Fe$_3$Sn$_2$}},\ }\bibfield  {journal}
	{\bibinfo  {journal} {Journal of Physics Condensed Matter}\ }\textbf
	{\bibinfo {volume} {21}},\ \href
	{https://doi.org/10.1088/0953-8984/21/45/452202}
	{10.1088/0953-8984/21/45/452202} (\bibinfo {year} {2009})\BibitemShut
	{NoStop}%
	\bibitem [{\citenamefont {Altthaler}\ \emph {et~al.}(2021)\citenamefont
		{Altthaler}, \citenamefont {Lysne}, \citenamefont {Roede}, \citenamefont
		{Prodan}, \citenamefont {Tsurkan}, \citenamefont {Kassem}, \citenamefont
		{Nakamura}, \citenamefont {Krohns}, \citenamefont {K{\'{e}}zsm{\'{a}}rki},\
		and\ \citenamefont {Meier}}]{Altthaler2021}%
	\BibitemOpen
	\bibfield  {author} {\bibinfo {author} {\bibfnamefont {M.}~\bibnamefont
			{Altthaler}}, \bibinfo {author} {\bibfnamefont {E.}~\bibnamefont {Lysne}},
		\bibinfo {author} {\bibfnamefont {E.}~\bibnamefont {Roede}}, \bibinfo
		{author} {\bibfnamefont {L.}~\bibnamefont {Prodan}}, \bibinfo {author}
		{\bibfnamefont {V.}~\bibnamefont {Tsurkan}}, \bibinfo {author} {\bibfnamefont
			{M.~A.}\ \bibnamefont {Kassem}}, \bibinfo {author} {\bibfnamefont
			{H.}~\bibnamefont {Nakamura}}, \bibinfo {author} {\bibfnamefont
			{S.}~\bibnamefont {Krohns}}, \bibinfo {author} {\bibfnamefont
			{I.}~\bibnamefont {K\'ezsm\'arki}},\ and\ \bibinfo {author}
		{\bibfnamefont {D.}~\bibnamefont {Meier}},\ }\bibfield  {title} {\bibinfo
		{title} {{Magnetic and geometric control of spin textures in the itinerant
				kagome magnet Fe$_3$Sn$_2$}},\ }\href
	{https://doi.org/10.1103/PhysRevResearch.3.043191} {\bibfield  {journal}
		{\bibinfo  {journal} {Physical Review Research}\ }\textbf {\bibinfo {volume}
			{3}},\ \bibinfo {pages} {043191} (\bibinfo {year} {2021})},\ \Eprint
	{https://arxiv.org/abs/2106.08791} {arXiv:2106.08791} \BibitemShut {NoStop}%
	\bibitem [{\citenamefont {Yao}\ \emph {et~al.}(2018)\citenamefont {Yao},
		\citenamefont {Lee}, \citenamefont {Xu}, \citenamefont {Wang}, \citenamefont
		{Ma}, \citenamefont {Yazyev}, \citenamefont {Xiong}, \citenamefont {Shi},
		\citenamefont {Aeppli},\ and\ \citenamefont {Soh}}]{Yao2018}%
	\BibitemOpen
	\bibfield  {author} {\bibinfo {author} {\bibfnamefont {M.}~\bibnamefont
			{Yao}}, \bibinfo {author} {\bibfnamefont {H.}~\bibnamefont {Lee}}, \bibinfo
		{author} {\bibfnamefont {N.}~\bibnamefont {Xu}}, \bibinfo {author}
		{\bibfnamefont {Y.}~\bibnamefont {Wang}}, \bibinfo {author} {\bibfnamefont
			{J.}~\bibnamefont {Ma}}, \bibinfo {author} {\bibfnamefont {O.~V.}\
			\bibnamefont {Yazyev}}, \bibinfo {author} {\bibfnamefont {Y.}~\bibnamefont
			{Xiong}}, \bibinfo {author} {\bibfnamefont {M.}~\bibnamefont {Shi}}, \bibinfo
		{author} {\bibfnamefont {G.}~\bibnamefont {Aeppli}},\ and\ \bibinfo {author}
		{\bibfnamefont {Y.}~\bibnamefont {Soh}},\ }\bibfield  {title} {\bibinfo
		{title} {{Switchable Weyl nodes in topological Kagome ferromagnet
				Fe$_3$Sn$_2$}},\ }\href {http://arxiv.org/abs/1810.01514} {\bibfield
		{journal} {\bibinfo  {journal} {arXiv}\ ,\ \bibinfo {pages} {1}} (\bibinfo
		{year} {2018})},\ \Eprint {https://arxiv.org/abs/1810.01514}
	{arXiv:1810.01514} \BibitemShut {NoStop}%
	\bibitem [{\citenamefont {Lin}\ and\ \citenamefont {Chen}(2020)}]{Lin2020}%
	\BibitemOpen
	\bibfield  {author} {\bibinfo {author} {\bibfnamefont {Z.~Z.}\ \bibnamefont
			{Lin}}\ and\ \bibinfo {author} {\bibfnamefont {X.}~\bibnamefont {Chen}},\
	}\bibfield  {title} {\bibinfo {title} {{Tunable Massive Dirac Fermions in
				Ferromagnetic Fe$_3$Sn$_2$ Kagome Lattice}},\ }\href
	{https://doi.org/10.1002/pssr.201900705} {\bibfield  {journal} {\bibinfo
			{journal} {Physica Status Solidi - Rapid Research Letters}\ }\textbf
		{\bibinfo {volume} {14}},\ \bibinfo {pages} {1} (\bibinfo {year}
		{2020})}\BibitemShut {NoStop}%
	\bibitem [{\citenamefont {Tanaka}\ \emph {et~al.}(2020)\citenamefont {Tanaka},
		\citenamefont {Fujisawa}, \citenamefont {Kuroda}, \citenamefont {Noguchi},
		\citenamefont {Sakuragi}, \citenamefont {Bareille}, \citenamefont {Smith},
		\citenamefont {Cacho}, \citenamefont {Jung}, \citenamefont {Muro},
		\citenamefont {Okada},\ and\ \citenamefont {Kondo}}]{Tanaka2020}%
	\BibitemOpen
	\bibfield  {author} {\bibinfo {author} {\bibfnamefont {H.}~\bibnamefont
			{Tanaka}}, \bibinfo {author} {\bibfnamefont {Y.}~\bibnamefont {Fujisawa}},
		\bibinfo {author} {\bibfnamefont {K.}~\bibnamefont {Kuroda}}, \bibinfo
		{author} {\bibfnamefont {R.}~\bibnamefont {Noguchi}}, \bibinfo {author}
		{\bibfnamefont {S.}~\bibnamefont {Sakuragi}}, \bibinfo {author}
		{\bibfnamefont {C.}~\bibnamefont {Bareille}}, \bibinfo {author}
		{\bibfnamefont {B.}~\bibnamefont {Smith}}, \bibinfo {author} {\bibfnamefont
			{C.}~\bibnamefont {Cacho}}, \bibinfo {author} {\bibfnamefont {S.~W.}\
			\bibnamefont {Jung}}, \bibinfo {author} {\bibfnamefont {T.}~\bibnamefont
			{Muro}}, \bibinfo {author} {\bibfnamefont {Y.}~\bibnamefont {Okada}},\ and\
		\bibinfo {author} {\bibfnamefont {T.}~\bibnamefont {Kondo}},\ }\bibfield
	{title} {\bibinfo {title} {{Three-dimensional electronic structure in
				ferromagnetic Fe$_3$Sn$_2$ with breathing kagome bilayers}},\ }\href
	{https://doi.org/10.1103/PhysRevB.101.161114} {\bibfield  {journal} {\bibinfo
			{journal} {Physical Review B}\ }\textbf {\bibinfo {volume} {101}},\ \bibinfo
		{pages} {161114(R)} (\bibinfo {year} {2020})}\BibitemShut {NoStop}%
	\bibitem [{\citenamefont {Fang}\ \emph {et~al.}(2022)\citenamefont {Fang},
		\citenamefont {Ye}, \citenamefont {Ghimire}, \citenamefont {Kang},
		\citenamefont {Liu}, \citenamefont {Han}, \citenamefont {Fu}, \citenamefont
		{Richter}, \citenamefont {van~den Brink}, \citenamefont {Kaxiras},
		\citenamefont {Comin},\ and\ \citenamefont {Checkelsky}}]{Fang2022}%
	\BibitemOpen
	\bibfield  {author} {\bibinfo {author} {\bibfnamefont {S.}~\bibnamefont
			{Fang}}, \bibinfo {author} {\bibfnamefont {L.}~\bibnamefont {Ye}}, \bibinfo
		{author} {\bibfnamefont {M.~P.}\ \bibnamefont {Ghimire}}, \bibinfo {author}
		{\bibfnamefont {M.}~\bibnamefont {Kang}}, \bibinfo {author} {\bibfnamefont
			{J.}~\bibnamefont {Liu}}, \bibinfo {author} {\bibfnamefont {M.}~\bibnamefont
			{Han}}, \bibinfo {author} {\bibfnamefont {L.}~\bibnamefont {Fu}}, \bibinfo
		{author} {\bibfnamefont {M.}~\bibnamefont {Richter}}, \bibinfo {author}
		{\bibfnamefont {J.}~\bibnamefont {van~den Brink}}, \bibinfo {author}
		{\bibfnamefont {E.}~\bibnamefont {Kaxiras}}, \bibinfo {author} {\bibfnamefont
			{R.}~\bibnamefont {Comin}},\ and\ \bibinfo {author} {\bibfnamefont {J.~G.}\
			\bibnamefont {Checkelsky}},\ }\bibfield  {title} {\bibinfo {title}
		{{Ferromagnetic helical nodal line and Kane-Mele spin-orbit coupling in
				kagome metal ${\mathrm{Fe}}_{3}{\mathrm{Sn}}_{2}$}},\ }\href
	{https://doi.org/10.1103/PhysRevB.105.035107} {\bibfield  {journal} {\bibinfo
			{journal} {Phys. Rev. B}\ }\textbf {\bibinfo {volume} {105}},\ \bibinfo
		{pages} {035107} (\bibinfo {year} {2022})}\BibitemShut {NoStop}%
	\bibitem [{\citenamefont {Lin}\ \emph {et~al.}(2018)\citenamefont {Lin},
		\citenamefont {Choi}, \citenamefont {Zhang}, \citenamefont {Qin},
		\citenamefont {Yi}, \citenamefont {Wang}, \citenamefont {Li}, \citenamefont
		{Wang}, \citenamefont {Zhang}, \citenamefont {Sun}, \citenamefont {Wei},
		\citenamefont {Zhang}, \citenamefont {Guo}, \citenamefont {Lu}, \citenamefont
		{Cho}, \citenamefont {Zeng},\ and\ \citenamefont {Zhang}}]{Lin2018}%
	\BibitemOpen
	\bibfield  {author} {\bibinfo {author} {\bibfnamefont {Z.}~\bibnamefont
			{Lin}}, \bibinfo {author} {\bibfnamefont {J.~H.}\ \bibnamefont {Choi}},
		\bibinfo {author} {\bibfnamefont {Q.}~\bibnamefont {Zhang}}, \bibinfo
		{author} {\bibfnamefont {W.}~\bibnamefont {Qin}}, \bibinfo {author}
		{\bibfnamefont {S.}~\bibnamefont {Yi}}, \bibinfo {author} {\bibfnamefont
			{P.}~\bibnamefont {Wang}}, \bibinfo {author} {\bibfnamefont {L.}~\bibnamefont
			{Li}}, \bibinfo {author} {\bibfnamefont {Y.}~\bibnamefont {Wang}}, \bibinfo
		{author} {\bibfnamefont {H.}~\bibnamefont {Zhang}}, \bibinfo {author}
		{\bibfnamefont {Z.}~\bibnamefont {Sun}}, \bibinfo {author} {\bibfnamefont
			{L.}~\bibnamefont {Wei}}, \bibinfo {author} {\bibfnamefont {S.}~\bibnamefont
			{Zhang}}, \bibinfo {author} {\bibfnamefont {T.}~\bibnamefont {Guo}}, \bibinfo
		{author} {\bibfnamefont {Q.}~\bibnamefont {Lu}}, \bibinfo {author}
		{\bibfnamefont {J.~H.}\ \bibnamefont {Cho}}, \bibinfo {author} {\bibfnamefont
			{C.}~\bibnamefont {Zeng}},\ and\ \bibinfo {author} {\bibfnamefont
			{Z.}~\bibnamefont {Zhang}},\ }\bibfield  {title} {\bibinfo {title}
		{{Flatbands and Emergent Ferromagnetic Ordering in Fe$_3$Sn$_2$ Kagome
				Lattices}},\ }\href {https://doi.org/10.1103/PhysRevLett.121.096401}
	{\bibfield  {journal} {\bibinfo  {journal} {Physical Review Letters}\
		}\textbf {\bibinfo {volume} {121}},\ \bibinfo {pages} {096401} (\bibinfo
		{year} {2018})}\BibitemShut {NoStop}%
	\bibitem [{\citenamefont {Bord{\'{a}}cs}\ \emph {et~al.}(2010)\citenamefont
		{Bord{\'{a}}cs}, \citenamefont {K{\'{e}}zsm{\'{a}}rki}, \citenamefont
		{Ohgushi},\ and\ \citenamefont {Tokura}}]{Bordacs2010}%
	\BibitemOpen
	\bibfield  {author} {\bibinfo {author} {\bibfnamefont {S.}~\bibnamefont
			{Bord{\'{a}}cs}}, \bibinfo {author} {\bibfnamefont {I.}~\bibnamefont
			{K{\'{e}}zsm{\'{a}}rki}}, \bibinfo {author} {\bibfnamefont {K.}~\bibnamefont
			{Ohgushi}},\ and\ \bibinfo {author} {\bibfnamefont {Y.}~\bibnamefont
			{Tokura}},\ }\bibfield  {title} {\bibinfo {title} {{Experimental band
				structure of the nearly half-metallic CuCr$_2$Se$_4$: An optical and
				magneto-optical study}},\ }\href
	{https://doi.org/10.1088/1367-2630/12/5/053039} {\bibfield  {journal}
		{\bibinfo  {journal} {New Journal of Physics}\ }\textbf {\bibinfo {volume}
			{12}},\ \bibinfo {pages} {0} (\bibinfo {year} {2010})},\ \Eprint
	{https://arxiv.org/abs/0907.5087} {arXiv:0907.5087} \BibitemShut {NoStop}%
	\bibitem [{\citenamefont {Shimano}\ \emph {et~al.}(2011)\citenamefont
		{Shimano}, \citenamefont {Ikebe}, \citenamefont {Takahashi}, \citenamefont
		{Kawasaki}, \citenamefont {Nagaosa},\ and\ \citenamefont
		{Tokura}}]{Shimano2011}%
	\BibitemOpen
	\bibfield  {author} {\bibinfo {author} {\bibfnamefont {R.}~\bibnamefont
			{Shimano}}, \bibinfo {author} {\bibfnamefont {Y.}~\bibnamefont {Ikebe}},
		\bibinfo {author} {\bibfnamefont {K.~S.}\ \bibnamefont {Takahashi}}, \bibinfo
		{author} {\bibfnamefont {M.}~\bibnamefont {Kawasaki}}, \bibinfo {author}
		{\bibfnamefont {N.}~\bibnamefont {Nagaosa}},\ and\ \bibinfo {author}
		{\bibfnamefont {Y.}~\bibnamefont {Tokura}},\ }\bibfield  {title} {\bibinfo
		{title} {Terahertz faraday rotation induced by an anomalous hall effect in
			the itinerant ferromagnet {SrRuO$_3$}},\ }\href
	{https://doi.org/10.1209/0295-5075/95/17002} {\bibfield  {journal} {\bibinfo
			{journal} {{EPL} (Europhysics Letters)}\ }\textbf {\bibinfo {volume} {95}},\
		\bibinfo {pages} {17002} (\bibinfo {year} {2011})}\BibitemShut {NoStop}%
	\bibitem [{\citenamefont {Okamura}\ \emph {et~al.}(2020)\citenamefont
		{Okamura}, \citenamefont {Minami}, \citenamefont {Kato}, \citenamefont
		{Fujishiro}, \citenamefont {Kaneko}, \citenamefont {Ikeda}, \citenamefont
		{Muramoto}, \citenamefont {Kaneko}, \citenamefont {Ueda}, \citenamefont
		{Kocsis}, \citenamefont {Kanazawa}, \citenamefont {Taguchi}, \citenamefont
		{Koretsune}, \citenamefont {Fujiwara}, \citenamefont {Tsukazaki},
		\citenamefont {Arita}, \citenamefont {Tokura},\ and\ \citenamefont
		{Takahashi}}]{Okamura2020}%
	\BibitemOpen
	\bibfield  {author} {\bibinfo {author} {\bibfnamefont {Y.}~\bibnamefont
			{Okamura}}, \bibinfo {author} {\bibfnamefont {S.}~\bibnamefont {Minami}},
		\bibinfo {author} {\bibfnamefont {Y.}~\bibnamefont {Kato}}, \bibinfo {author}
		{\bibfnamefont {Y.}~\bibnamefont {Fujishiro}}, \bibinfo {author}
		{\bibfnamefont {Y.}~\bibnamefont {Kaneko}}, \bibinfo {author} {\bibfnamefont
			{J.}~\bibnamefont {Ikeda}}, \bibinfo {author} {\bibfnamefont
			{J.}~\bibnamefont {Muramoto}}, \bibinfo {author} {\bibfnamefont
			{R.}~\bibnamefont {Kaneko}}, \bibinfo {author} {\bibfnamefont
			{K.}~\bibnamefont {Ueda}}, \bibinfo {author} {\bibfnamefont {V.}~\bibnamefont
			{Kocsis}}, \bibinfo {author} {\bibfnamefont {N.}~\bibnamefont {Kanazawa}},
		\bibinfo {author} {\bibfnamefont {Y.}~\bibnamefont {Taguchi}}, \bibinfo
		{author} {\bibfnamefont {T.}~\bibnamefont {Koretsune}}, \bibinfo {author}
		{\bibfnamefont {K.}~\bibnamefont {Fujiwara}}, \bibinfo {author}
		{\bibfnamefont {A.}~\bibnamefont {Tsukazaki}}, \bibinfo {author}
		{\bibfnamefont {R.}~\bibnamefont {Arita}}, \bibinfo {author} {\bibfnamefont
			{Y.}~\bibnamefont {Tokura}},\ and\ \bibinfo {author} {\bibfnamefont
			{Y.}~\bibnamefont {Takahashi}},\ }\bibfield  {title} {\bibinfo {title}
		{{Giant magneto-optical responses in magnetic Weyl semimetal
				Co$_3$Sn$_2$S$_2$}},\ }\bibfield  {journal} {\bibinfo  {journal} {Nature
			Communications}\ }\textbf {\bibinfo {volume} {11}},\ \href
	{https://doi.org/10.1038/s41467-020-18470-0} {10.1038/s41467-020-18470-0}
	(\bibinfo {year} {2020})\BibitemShut {NoStop}%
	\bibitem [{\citenamefont {Jones}\ and\ \citenamefont
		{Gunnarsson}(1989)}]{jo.gu.89}%
	\BibitemOpen
	\bibfield  {author} {\bibinfo {author} {\bibfnamefont {R.~O.}\ \bibnamefont
			{Jones}}\ and\ \bibinfo {author} {\bibfnamefont {O.}~\bibnamefont
			{Gunnarsson}},\ }\bibfield  {title} {\bibinfo {title} {The density functional
			formalism, its applications and prospects},\ }\href
	{https://doi.org/10.1103/RevModPhys.61.689} {\bibfield  {journal} {\bibinfo
			{journal} {Rev. Mod. Phys.}\ }\textbf {\bibinfo {volume} {61}},\ \bibinfo
		{pages} {689} (\bibinfo {year} {1989})}\BibitemShut {NoStop}%
	\bibitem [{\citenamefont {Kohn}(1999)}]{kohn.99}%
	\BibitemOpen
	\bibfield  {author} {\bibinfo {author} {\bibfnamefont {W.}~\bibnamefont
			{Kohn}},\ }\bibfield  {title} {\bibinfo {title} {Nobel lecture: Electronic
			structure of matter\char22{}wave functions and density functionals},\
	}\href@noop {} {\bibfield  {journal} {\bibinfo  {journal} {Rev. Mod. Phys.}\
		}\textbf {\bibinfo {volume} {71}},\ \bibinfo {pages} {1253} (\bibinfo {year}
		{1999})}\BibitemShut {NoStop}%
	\bibitem [{\citenamefont {Jones}(2015)}]{jone.15}%
	\BibitemOpen
	\bibfield  {author} {\bibinfo {author} {\bibfnamefont {R.~O.}\ \bibnamefont
			{Jones}},\ }\bibfield  {title} {\bibinfo {title} {Density functional theory:
			Its origins, rise to prominence, and future},\ }\href
	{https://doi.org/10.1103/RevModPhys.87.897} {\bibfield  {journal} {\bibinfo
			{journal} {Rev. Mod. Phys.}\ }\textbf {\bibinfo {volume} {87}},\ \bibinfo
		{pages} {897} (\bibinfo {year} {2015})}\BibitemShut {NoStop}%
	\bibitem [{\citenamefont {Hasan}\ and\ \citenamefont {Kane}(2010)}]{ha.ka.10}%
	\BibitemOpen
	\bibfield  {author} {\bibinfo {author} {\bibfnamefont {M.~Z.}\ \bibnamefont
			{Hasan}}\ and\ \bibinfo {author} {\bibfnamefont {C.~L.}\ \bibnamefont
			{Kane}},\ }\bibfield  {title} {\bibinfo {title} {Colloquium: Topological
			insulators},\ }\href {https://doi.org/10.1103/RevModPhys.82.3045} {\bibfield
		{journal} {\bibinfo  {journal} {Rev. Mod. Phys.}\ }\textbf {\bibinfo {volume}
			{82}},\ \bibinfo {pages} {3045} (\bibinfo {year} {2010})}\BibitemShut
	{NoStop}%
	\bibitem [{\citenamefont {Xiao}\ \emph {et~al.}(2010)\citenamefont {Xiao},
		\citenamefont {Chang},\ and\ \citenamefont {Niu}}]{xi.mi.10}%
	\BibitemOpen
	\bibfield  {author} {\bibinfo {author} {\bibfnamefont {D.}~\bibnamefont
			{Xiao}}, \bibinfo {author} {\bibfnamefont {M.-C.}\ \bibnamefont {Chang}},\
		and\ \bibinfo {author} {\bibfnamefont {Q.}~\bibnamefont {Niu}},\ }\bibfield
	{title} {\bibinfo {title} {Berry phase effects on electronic properties},\
	}\href {https://doi.org/10.1103/RevModPhys.82.1959} {\bibfield  {journal}
		{\bibinfo  {journal} {Rev. Mod. Phys.}\ }\textbf {\bibinfo {volume} {82}},\
		\bibinfo {pages} {1959} (\bibinfo {year} {2010})}\BibitemShut {NoStop}%
	\bibitem [{\citenamefont {Feil}\ and\ \citenamefont {Haas}(1987)}]{Feil1987}%
	\BibitemOpen
	\bibfield  {author} {\bibinfo {author} {\bibfnamefont {H.}~\bibnamefont
			{Feil}}\ and\ \bibinfo {author} {\bibfnamefont {C.}~\bibnamefont {Haas}},\
	}\bibfield  {title} {\bibinfo {title} {{Magneto-Optical Kerr Effect, Enhanced
				by the Plasma Resonance of Charge Carriers}},\ }\href
	{https://doi.org/10.1038/182227e0} {\bibfield  {journal} {\bibinfo  {journal}
			{Physical Review Letters}\ }\textbf {\bibinfo {volume} {58}},\ \bibinfo
		{pages} {65} (\bibinfo {year} {1987})}\BibitemShut {NoStop}%
	\bibitem [{Note1()}]{Note1}%
	\BibitemOpen
	\bibinfo {note} {The experimental details together with the dc AHE data and
		details of the crystal growth are presented in the Supplemental
		Material.}\BibitemShut {Stop}%
	\bibitem [{\citenamefont {Blaha}\ \emph {et~al.}(2018)\citenamefont {Blaha},
		\citenamefont {Schwarz}, \citenamefont {Madsen}, \citenamefont {Kvasnicka},
		\citenamefont {Luitz}, \citenamefont {Laskowski}, \citenamefont {Tran},\ and\
		\citenamefont {Marks.}}]{Blaha2021}%
	\BibitemOpen
	\bibfield  {author} {\bibinfo {author} {\bibfnamefont {P.}~\bibnamefont
			{Blaha}}, \bibinfo {author} {\bibfnamefont {K.}~\bibnamefont {Schwarz}},
		\bibinfo {author} {\bibfnamefont {G.~K.~H.}\ \bibnamefont {Madsen}}, \bibinfo
		{author} {\bibfnamefont {D.}~\bibnamefont {Kvasnicka}}, \bibinfo {author}
		{\bibfnamefont {J.}~\bibnamefont {Luitz}}, \bibinfo {author} {\bibfnamefont
			{R.}~\bibnamefont {Laskowski}}, \bibinfo {author} {\bibfnamefont
			{F.}~\bibnamefont {Tran}},\ and\ \bibinfo {author} {\bibfnamefont {L.~D.}\
			\bibnamefont {Marks.}},\ }\href@noop {} {\emph {\bibinfo {title} {{WIEN2k, An
					Augmented Plane Wave + Local Orbitals Program for Calculating Crystal
					Properties}}}}\ (\bibinfo  {publisher} {Karlheinz Schwarz, Techn.
		Universität Wien, Austria},\ \bibinfo {year} {2018})\BibitemShut {NoStop}%
	\bibitem [{\citenamefont {Perdew}\ \emph {et~al.}(1996)\citenamefont {Perdew},
		\citenamefont {Burke},\ and\ \citenamefont {Ernzerhof}}]{pe.bu.96}%
	\BibitemOpen
	\bibfield  {author} {\bibinfo {author} {\bibfnamefont {J.~P.}\ \bibnamefont
			{Perdew}}, \bibinfo {author} {\bibfnamefont {K.}~\bibnamefont {Burke}},\ and\
		\bibinfo {author} {\bibfnamefont {M.}~\bibnamefont {Ernzerhof}},\ }\bibfield
	{title} {\bibinfo {title} {Generalized gradient approximation made simple},\
	}\href {https://doi.org/10.1103/PhysRevLett.77.3865} {\bibfield  {journal}
		{\bibinfo  {journal} {Phys. Rev. Lett.}\ }\textbf {\bibinfo {volume} {77}},\
		\bibinfo {pages} {3865} (\bibinfo {year} {1996})}\BibitemShut {NoStop}%
	\bibitem [{Note2()}]{Note2}%
	\BibitemOpen
	\bibinfo {note} {Https://www.youtube.com/watch?v=dWJxK6io0UA, in private
		communication, the presenter confirmed that these results have entered the
		publication process. Once available the reference will be
		updated.}\BibitemShut {Stop}%
	\bibitem [{\citenamefont {Kubo}(1957)}]{kubo.57}%
	\BibitemOpen
	\bibfield  {author} {\bibinfo {author} {\bibfnamefont {R.}~\bibnamefont
			{Kubo}},\ }\bibfield  {title} {\bibinfo {title} {Statistical-mechanical
			theory of irreversible processes. i. general theory and simple applications
			to magnetic and conduction problems},\ }\href
	{https://doi.org/10.1143/JPSJ.12.570} {\bibfield  {journal} {\bibinfo
			{journal} {Journal of the Physical Society of Japan}\ }\textbf {\bibinfo
			{volume} {12}},\ \bibinfo {pages} {570} (\bibinfo {year} {1957})},\ \Eprint
	{https://arxiv.org/abs/https://doi.org/10.1143/JPSJ.12.570}
	{https://doi.org/10.1143/JPSJ.12.570} \BibitemShut {NoStop}%
	\bibitem [{\citenamefont {Tanner}(2015)}]{Tanner.2015}%
	\BibitemOpen
	\bibfield  {author} {\bibinfo {author} {\bibfnamefont {D.~B.}\ \bibnamefont
			{Tanner}},\ }\bibfield  {title} {\bibinfo {title} {Use of x-ray scattering
			functions in kramers-kronig analysis of reflectance},\ }\href
	{https://doi.org/10.1103/PhysRevB.91.035123} {\bibfield  {journal} {\bibinfo
			{journal} {Phys. Rev. B}\ }\textbf {\bibinfo {volume} {91}},\ \bibinfo
		{pages} {035123} (\bibinfo {year} {2015})}\BibitemShut {NoStop}%
	\bibitem [{\citenamefont {Sato}(1981)}]{Sato1981}%
	\BibitemOpen
	\bibfield  {author} {\bibinfo {author} {\bibfnamefont {K.}~\bibnamefont
			{Sato}},\ }\bibfield  {title} {\bibinfo {title} {{Measurement of
				Magneto-Optical Kerr Effect Using Piezo-Birefringent Modulator}},\
	}\href@noop {} {\bibfield  {journal} {\bibinfo  {journal} {Japanese Journal
				of Applied Physics}\ }\textbf {\bibinfo {volume} {20}},\ \bibinfo {pages}
		{2403} (\bibinfo {year} {1981})}\BibitemShut {NoStop}%
	\bibitem [{\citenamefont {Demk{\'{o}}}\ \emph {et~al.}(2012)\citenamefont
		{Demk{\'{o}}}, \citenamefont {Schober}, \citenamefont {Kocsis}, \citenamefont
		{Bahramy}, \citenamefont {Murakawa}, \citenamefont {Lee}, \citenamefont
		{K{\'{e}}zsm{\'{a}}rki}, \citenamefont {Arita}, \citenamefont {Nagaosa},\
		and\ \citenamefont {Tokura}}]{Demko2012}%
	\BibitemOpen
	\bibfield  {author} {\bibinfo {author} {\bibfnamefont {L.}~\bibnamefont
			{Demk{\'{o}}}}, \bibinfo {author} {\bibfnamefont {G.~A.}\ \bibnamefont
			{Schober}}, \bibinfo {author} {\bibfnamefont {V.}~\bibnamefont {Kocsis}},
		\bibinfo {author} {\bibfnamefont {M.~S.}\ \bibnamefont {Bahramy}}, \bibinfo
		{author} {\bibfnamefont {H.}~\bibnamefont {Murakawa}}, \bibinfo {author}
		{\bibfnamefont {J.~S.}\ \bibnamefont {Lee}}, \bibinfo {author} {\bibfnamefont
			{I.}~\bibnamefont {K{\'{e}}zsm{\'{a}}rki}}, \bibinfo {author} {\bibfnamefont
			{R.}~\bibnamefont {Arita}}, \bibinfo {author} {\bibfnamefont
			{N.}~\bibnamefont {Nagaosa}},\ and\ \bibinfo {author} {\bibfnamefont
			{Y.}~\bibnamefont {Tokura}},\ }\bibfield  {title} {\bibinfo {title}
		{{Enhanced infrared magneto-optical response of the nonmagnetic semiconductor
				BiTeI driven by bulk rashba splitting}},\ }\href
	{https://doi.org/10.1103/PhysRevLett.109.167401} {\bibfield  {journal}
		{\bibinfo  {journal} {Physical Review Letters}\ }\textbf {\bibinfo {volume}
			{109}},\ \bibinfo {pages} {1} (\bibinfo {year} {2012})}\BibitemShut {NoStop}%
	
	\bibitem [{\citenamefont {Kuzmenko}(2005)}]{Kuzmenko2005}%
	\BibitemOpen
	\bibfield  {author} {\bibinfo {author} {\bibfnamefont {A.~B.}\ \bibnamefont
			{Kuzmenko}},\ }\bibfield  {title} {\bibinfo {title} {{Kramers-Kronig
				constrained variational analysis of optical spectra}},\ }\href
	{https://doi.org/10.1063/1.1979470} {\bibfield  {journal} {\bibinfo
			{journal} {Review of Scientific Instruments}\ }\textbf {\bibinfo {volume}
			{76}},\ \bibinfo {pages} {1} (\bibinfo {year} {2005})},\ \Eprint
	{https://arxiv.org/abs/0503565} {arXiv:0503565 [cond-mat]} \BibitemShut
	{NoStop}%
	\bibitem [{\citenamefont {Ambrosch-Draxl}\ and\ \citenamefont
		{Sofo}(2006)}]{ambrosch-draxl_linear_2006}%
	\BibitemOpen
	\bibfield  {author} {\bibinfo {author} {\bibfnamefont {C.}~\bibnamefont
			{Ambrosch-Draxl}}\ and\ \bibinfo {author} {\bibfnamefont {J.~O.}\
			\bibnamefont {Sofo}},\ }\bibfield  {title} {\bibinfo {title} {Linear optical
			properties of solids within the full-potential linearized augmented planewave
			method},\ }\href {https://doi.org/10.1016/j.cpc.2006.03.005} {\bibfield
		{journal} {\bibinfo  {journal} {Computer Physics Communications}\ }\textbf
		{\bibinfo {volume} {175}},\ \bibinfo {pages} {1} (\bibinfo {year}
		{2006})}\BibitemShut {NoStop}%
\end{thebibliography}
\end{document}